\documentclass[preprint,pre,showpacs,preprintnumbers,amsmath,amssymb]{revtex4}

\usepackage{graphicx}
\usepackage{dcolumn}
\usepackage{bm}
\usepackage{braket}

\begin{document}

\preprint{PREPRINT}

\title{Incorporating Forcing Terms in Cascaded Lattice-Boltzmann Approach by Method of Central Moments}

\author{Kannan N. Premnath}
\email{nandha@metah.com}

\affiliation{Department of Chemical Engineering, University of
California, Santa Barbara, Santa Barbara, CA 93106\\}

\affiliation{ MetaHeuristics LLC, 3944 State Street, Suite 350,
Santa Barbara, CA 93105}

\author{Sanjoy Banerjee}
\email{banerjee@engineering.ucsb.edu}

\affiliation{Department of Chemical Engineering\\ Department of
Mechanical Engineering\\ Bren School of Environmental Science and Management\\
University of California, Santa Barbara, Santa Barbara, CA 93106\\}

\date{\today}

\begin{abstract}
Cascaded lattice-Boltzmann method (Cascaded-LBM) employs a new class of collision operators aiming to stabilize computations and remove certain modeling artifacts for simulation of fluid flow on lattice grids with sizes arbitrarily larger than the smallest physical dissipation length scale (Geier \emph{et al.}, Phys. Rev. E $\mathbf{63}$, 066705 (2006)). It achieves this and distinguishes from other collision operators, such as in the standard single or multiple relaxation time approaches, by performing relaxation process due to collisions in terms of moments shifted by the local hydrodynamic fluid velocity, i.e. central moments, in an ascending order-by-order at different relaxation rates. In this paper, we propose and derive source terms in the Cascaded-LBM to represent the effect of external or internal forces on the dynamics of fluid motion. This is essentially achieved by matching the continuous form of the central moments of the source or forcing terms with its discrete version. Different forms of continuous central moments of sources, including one that is obtained from a local Maxwellian, are considered in this regard. As a result, the forcing terms obtained in this new formulation are Galilean invariant by construction. To alleviate lattice artifacts due to forcing terms
in the emergent macroscopic fluid equations, they are proposed as temporally semi-implicit and second-order, and the implicitness
is subsequently effectively removed by means of a transformation to facilitate computation. It is shown that the impressed force field influences the cascaded collision process in the evolution of the transformed distribution function. The method of central moments along with the associated orthogonal properties of the moment basis completely determines the analytical expressions for the source terms as a function of the force and macroscopic velocity fields. In contrast to the existing forcing schemes, it is found that they involve higher order terms in velocity space. It is shown that the proposed approach implies ``generalization'' of both local equilibrium and source terms in the usual lattice frame of reference, which depend on the ratio of the relaxation times of moments of different orders. An analysis by means of the Chapman-Enskog multiscale expansion shows that the Cascaded-LBM with forcing terms is consistent with the Navier-Stokes equations. Computational experiments with canonical problems involving different types of forces demonstrate its accuracy.
\end{abstract}

\pacs{47.11.Qr,05.20.Dd,47.27.-i}
\maketitle

\section{\label{sec:intro}Introduction}
Lattice-Boltzmann method (LBM), based on minimal discrete kinetic models, has attracted considerable attention as an alternative computational approach for fluid mechanics problems~\cite{benzi92,chen98,succi01,yu03}. While its origins can be traced to lattice gas automata~\cite{frisch86} as a means to remove its statistical noise~\cite{mcnamara88}, over the years, the LBM has undergone major series of advances to improve its underlying models for better physical fidelity and computational efficiency. Moreover, its connection to the continuous Boltzmann equation as a dramatically simplified version~\cite{he97b,shan98} established it as an efficient approach in computational kinetic theory and led to the development of asymptotic tools~\cite{junk05} providing a rigorous framework for numerical consistency analysis. The LBM is based on performing stream-and-collide steps to compute the evolution of the distribution of particle populations, such that its averaged behavior recovers the dynamics of fluid motion. The streaming step is a free-flight process along discrete characteristic particle directions designed from symmetry considerations, while the collision step is generally represented as a relaxation process of the distribution function to its attractors, i.e. local equilibrium states. Considerable effort has been made in developing models to account for various aspects of the collision process, as it has paramount influence on the physical fidelity and numerical stability of the LBM.

One of the simplest and among the most common is the single-relaxation-time (SRT) model proposed by
Chen \emph{et al}.~\cite{chen92} and Qian \emph{et al}.~\cite{qian92}, which is based on the BGK approximation~\cite{bhatnagar54}. On the other hand, d'Humi{\`e}res (1992)~\cite{dhumieres92} proposed a moment method, in which various moments that are integral properties of distribution functions weighted by the Cartesian components of discrete particle velocities of various orders are relaxed to their equilibrium states at different rates during collision step, leading to the multiple-relaxation-time (MRT) model. It is an important extension of the relaxation LBM proposed earlier by Higuera \emph{et al}~\cite{higuera89a,higuera89b}. While it is a much simplified version of the latter, the major innovation lies in representing the collision process in moment space~\cite{grad49} rather than the usual particle velocity space. By carefully separating the relaxation times of hydrodynamic and non-hydrodynamic moments, it has been shown that the MRT-LBM significantly improves the numerical stability~\cite{lallemand00,dhumieres02} and better physical representation in certain problems such as kinetic layers near boundaries~\cite{ginzburg03}, when compared with the SRT-LBM. Such MRT models have recently been shown to reproduce challenging fluid mechanics problems such as complex turbulent flows with good quantitative accuracy~\cite{premnath09,premnath09a}. An important and natural simplification of the MRT model is the two-relaxation-time (TRT) model, in which the moments of even and odd orders are relaxed at different rates~\cite{ginzburg05}.

From a different perspective, Karlin and co-workers~\cite{karlin99,boghosian01,ansumali02,succi02,karlin06} have developed the so-called entropic LBM in which the collision process is modeled by assuming that distribution functions are drawn towards their attractors, which are obtained by the minimization of a Lyapanov-type functional, i.e. the so-called H-theorem is enforced locally, while modulating the relaxation process with a single relaxation time to maintain numerical stability. It may be noted that in contrast to the SRT or MRT collision operators, which employ equilibria that are polynomials in hydrodynamic fields, the entropic collision operator, in general, requires use of non-polynomial or transcendental functions of hydrodynamic fields. Recently, using
this framework, a novel entropy-based MRT model was derived~\cite{asinari09} and a Galilean invariance restoration approach was
developed~\cite{prasianakis09}. In addition, there has been considerable progress in the development of systematic procedures
for high-order lattice-Boltzmann models~\cite{shan06,chikatamarla06}.

Recently, Geier \emph{et al}.~\cite{geier06} introduced another novel class of collision operator leading to the so-called Cascaded-LBM. Collision operators, such as the standard SRT or MRT models, are generally constructed to recover the Navier-Stokes equations (NSE), with errors that are quadratic in fluid velocity. Such models, which are Galilean invariant up to a lower degree, i.e., the square of Mach number, are prone to numerical instability, which can be alleviated to a degree with the use of the latter model. Recognizing that insufficient level of Galilean invariance is one of the main sources of numerical instability, Geier proposed to perform collision process in a frame of reference shifted by the macroscopic fluid velocity. Unlike other collision operators which perform relaxation in a special rest or lattice frame of reference, Cascaded-LBM chooses an intrinsic frame of reference obtained from the properties of the system itself. The local hydrodynamic velocity, which is the first moment of the distribution functions, is the center of mass in the space of moments. A coordinate system moving locally with this velocity at each node is a natural framework to describe the physics of collisions in the space of moments. This could enable achieving a higher degree of Galilean invariance than possible with the prior approaches. It may be noted that the moments displaced by the local hydrodynamic velocity are termed as the \emph{central moments} and are computed in a moving frame of reference. On the other hand, the moments with no such shift are called the \emph{raw moments}, which are computed in a rest frame of reference.

Based on this insight, the collision operator is constructed in such a way that each central moment can be relaxed independently with generally different relaxation rates. However, it is computationally easier to perform operations in terms of raw moments. Both forms of moments can be related to one another in terms of the binomial theorem, and hence the latter plays an important role in the construction of an operational collision step. As a result of this theorem, central moment of a given order are algebraic combinations of raw moments of different orders, with their highest order being equal to that of the central moment. In effect, the evolution of lower order raw moments influences higher order central moments and not vice versa. Thus, due to this specific directionality of coupling between different central and raw moments, starting from the lowest central moment, we can relax successively higher order central moments towards their equilibrium, which are implicitly carried out in terms of raw moments. Such structured sequential computation of relaxation in an ascending order of moments leads to a novel cascaded collision operator, in which the post-collision moments depend not only on the conserved moments, but also on the non-conserved moments and on each other.

Moreover, it was found that relaxing different central moments differently, certain artifacts such as aliasing that cause numerical instability for computation on coarse grids, whose sizes can be arbitrarily larger than the smallest physical or viscous dissipation length scale can be avoided. In particular, this is achieved by setting the third-order central moments to its equilibrium value, while allowing only the second-order moments to undergo over-relaxtion~\cite{geier08a}. The limit of stability is now dictated only by the Courant-Friedrichs-Lewy condition~\cite{courant28} typical of explicit schemes and not by effects arising due to the discreteness
of the particle velocity set. Prevention of such ultra-violet catastrophe in under-resolved computations could enable application of the LBM for high Reynolds number flows or for fluid with low viscosities. Further insight into the nature of the gain in numerical stability with Cascaded-LBM is achieved with the recognition that unlike other collision operators which appear to introduce de-stabilizing negative hyper-viscosity effects that are of second-order in Mach number due to insufficient Galilean invariance, the former seems to have stabilizing positive and smaller hyper-viscosity effects that are of fourth-order in Mach number~\cite{geier08b}. Recently, Asinari~\cite{asinari08} showed that cascaded relaxation using multiple relaxation times
is equivalent to performing relaxation to a ``generalized'' local equilibrium in the rest frame of reference. Such generalized local equilibrium is dependent on non-conserved moments as well as the ratio of various relaxation times.

Clearly, several situations exist in which the dynamics of fluid motion is driven or affected by the presence of external or self-consistent internal forces. Examples include gravity, magnetohydrodynamic forces, self-consistent internal forces
in multi-phase or multi-fluid systems. Moreover, subgrid scale (SGS) models for turbulence simulation can be explicitly introduced as body forces in kinetic approaches~\cite{girimaji07,premnath09a}. Thus, it is important to develop a consistent approach to introduce the effect of forces that act on the fluid flow in the Cascaded-LBM. The method for introducing force terms in other LBM approaches are given, for example, in ~\cite{he98,martys98,ladd01,guo02}, in which notably Guo~\emph{et al}.~\cite{guo02} developed a consistent approach which avoided spurious effects in the macroscopic equations resulting from the finiteness of the lattice set.

The approach proposed in this paper consists as follows. It consists of deriving forcing terms which can be obtained by matching their discrete central moments to their corresponding continuous version. In this regard, we consider two different sets of ansatz for the continuous source central moments -- one based on a continuous local Maxwellian and another one which makes specific assumptions regarding the effect of forces for higher order moments. An important feature of our approach is that by construction the source
terms are Galilean invariant, which would be a very desirable aspect from both physical and computational points of view.
To facilitate computation, the central source moments are related to corresponding raw moments, which are, in turn, expressed in velocity space. Furthermore, to improve temporal accuracy, the source terms are treated semi-implicitly. The implicitness, then, is effectively removed by applying a transformation to the distribution function. A detailed \emph{a priori} derivation of this central moment method is given so that it provides a mathematical framework which could also be useful for extension to other problems.
We then establish the consistency of our approach to macroscopic fluid dynamical equations by performing a Chapman-Enskog multiscale moment expansion. It will be shown that when Cascaded-LBM with forcing terms is reinterpreted in terms of the rest frame of reference (as usual with other LBM), it implies considering a generalized local equilibrium and sources, which also depend on the ratio of the relaxation times of various moments, for their higher order moments. Numerical experiments will also be performed to confirm the accuracy of our approach for flows with different types of forces, where analytical solutions are available.

This paper is structured as follows. Section~\ref{sec:discreteparticlevelocity} briefly discusses the choice of moment basis employed in this paper. In Sec.~\ref{sec:ccentraleqmmoments}, continuous forms of central moments for equilibrium and sources (for a specific ansatz) are introduced. The Cascaded-LBE with forcing terms are presented in Sec.~\ref{sec:cascadedLBEforcing}. In Sec.~\ref{sec:cascadedcollisionforcing}, we discuss the details of an analysis and the construction of the Cascaded-LBM and the analytical expressions for source terms. Section~\ref{sec:dealiasinghighercentralsourcemoments} provides the details of how the
computational procedure is modified with the use of a different form of the central source moments.
The computational procedure for Cascaded-LBM with forcing is provided in Sec.~\ref{sec:computationalprocedure}.
Results of the computational procedure for some canonical problems are presented in Sec.~\ref{sec:compexperiments}.
Summary and conclusions of this work are described in Sec.~\ref{sec:summaryconclusions}. Consistency analysis of the
central moment method with forcing terms by means of a Chapman-Enskog multiscale moment expansion is presented in Appendix~\ref{app:ChapmanEnskoganalysis}. Appendix~\ref{app:GeneralizedEqilibriumSources} shows that Cascaded-LBM with forcing terms is equivalent to considering a generalized local equilibrium and sources in the rest frame of reference.
Finally, Appendix~\ref{app:TimeImplicitnessCascadedCollision} investigates the possibility of introducing time-implicitness in the cascaded collision operator.

\section{\label{sec:discreteparticlevelocity}Choice of Basis Vectors for Moments}
\begin{figure}
\includegraphics[width = 65mm]{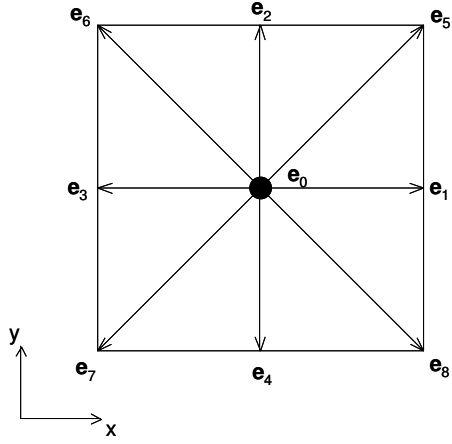}
\caption{\label{fig:d2q9} Two-dimensional, nine-velocity (D2Q9) Lattice.}
\end{figure}

For concreteness, without losing generality, we consider, the two-dimensional, nine velocity (D2Q9) model, which is shown in Fig.~\ref{fig:d2q9}. The particle velocity $\overrightarrow{e}_{\alpha}$ may be written as
\begin{equation}
\overrightarrow{e_{\alpha}} = \left\{\begin{array}{ll}
   {(0,0)}&{\alpha=0}\\
   {(\pm 1,0),(0,\pm 1)}&{\alpha=1,\cdots,4}\\
   {(\pm 1,\pm 1)} &{\alpha=5,\cdots,8}
\end{array} \right.
\label{eq:velocityd2q9}
\end{equation}
Here and henceforth, we employ Greek and Latin subscripts for particle velocity directions and Cartesian coordinate directions, respectively. Moments in the LBM are discrete integral properties of the distribution function $f_{\alpha}$, i.e. $\sum_{\alpha=0}^8e_{\alpha x}^m e_{\alpha y}^nf_{\alpha}$, where $m$ and $n$ are integers. Since the theory of the moment method draws heavily upon the associated orthogonality properties, for convenience, we employ the Dirac's bra-ket notation in this paper. That is, we denote the ``bra" operator $\bra{\phi}$ to represent a row vector of any state variable $\phi$ along each of the particle directions, i.e. $(\phi_0,\phi_1,\phi_2,\ldots,\phi_8)$, and the ``ket" operator $\ket{\phi}$ represents a column vector, i.e. $(\phi_0,\phi_1,\phi_2,\ldots,\phi_8)^\dag$, where the superscript ``\dag" is the transpose operator. In this notation, $\braket{\phi|\varphi}$ represents the inner-product, i.e. $\sum_{\alpha=0}^8\phi_{\alpha}\varphi_{\alpha}$. To obtain a moment space of the distribution functions, we start with a set of the following nine non-orthogonal basis vectors obtained from the combinations of the monomials $e_{\alpha x}^m e_{\alpha y}^n$ in an ascending order.
\begin{eqnarray}
\ket{\rho}\equiv \ket{|\overrightarrow{e}_{\alpha}|^{0}} &=&\left(1,1,1,1,1,1,1,1,1\right)^\dag, \\
\ket{e_{\alpha x}} &=&\left(0,1,0,-1,0,1,-1,-1,1\right)^\dag, \\
\ket{e_{\alpha y}} &=&\left(0,0,1,0,-1,1,1,-1,-1\right)^\dag,\\
\ket{e_{\alpha x}^2+e_{\alpha y}^2} &=&\left(0,1,1,1,1,2,2,2,2\right)^\dag,\\
\ket{e_{\alpha x}^2-e_{\alpha y}^2} &=&\left(0,1,-1,1,-1,0,0,0,0\right)^\dag,\\
\ket{e_{\alpha x}e_{\alpha y}} &=&\left(0,0,0,0,0,1,-1,1,-1\right)^\dag,\\
\ket{e_{\alpha x}^2e_{\alpha y}} &=&\left(0,0,0,0,0,1,1,-1,-1\right)^\dag,\\
\ket{e_{\alpha x}e_{\alpha y}^2} &=&\left(0,0,0,0,0,1,-1,-1,1\right)^\dag,\\
\ket{e_{\alpha x}^2e_{\alpha y}^2} &=&\left(0,0,0,0,0,1,1,1,1\right)^\dag.
\end{eqnarray}

To facilitate analysis, the above set of basis vectors is transformed into an equivalent \emph{orthogonal} set of basis vectors by means of the standard Gram-Schmidt procedure in the increasing order of the monomials of the products of the Cartesian components of the particle velocities:
\begin{eqnarray}
\ket{K_{0}}&=&\ket{\rho},\\
\ket{K_{1}}&=&\ket{e_{\alpha x}},\\
\ket{K_{2}}&=&\ket{e_{\alpha y}},\\
\ket{K_{3}}&=&3\ket{e_{\alpha x}^2+e_{\alpha y}^2}-4\ket{\rho},\\
\ket{K_{4}}&=&\ket{e_{\alpha x}^2-e_{\alpha y}^2},\\
\ket{K_{5}}&=&\ket{e_{\alpha x}e_{\alpha y}},\\
\ket{K_{6}}&=&-3\ket{e_{\alpha x}^2e_{\alpha y}}+2\ket{e_{\alpha y}},\\
\ket{K_{7}}&=&-3\ket{e_{\alpha x}e_{\alpha y}^2}+2\ket{e_{\alpha x}},\\
\ket{K_{8}}&=&9\ket{e_{\alpha x}^2e_{\alpha y}^2}-6\ket{e_{\alpha x}^2+e_{\alpha y}^2}+4\ket{\rho}.
\end{eqnarray}
This is very similar to that used by Geier \emph{et al}.~\cite{geier06}, except for the negative sign used in $\ket{K_{5}}$ by the latter. The purpose of using a slightly different orthogonal basis than that considered in~\cite{geier06} is simply to illustrate how it changes the details of the cascaded collision operator. It is obvious that we can define different sets of orthogonal basis vectors that differ from one another by a constant factor or a sign. Furthermore, it is noteworthy to compare
the ordering of basis vectors used for the central moment method with that considered by Lallemand and Luo~\cite{lallemand00}: Here, the ordering is based on the ascending powers of moments (i.e. zeroth order moment, first order moments, second order moments,$\ldots$) while~\cite{lallemand00} order their basis vectors based on the character of moments, i.e. increasing powers of their tensorial orders (i.e. scalars, vectors, tensors of different ranks,$\ldots$).

The orthogonal set of basis vectors can be written in terms of the following matrix
\begin{equation}
\mathcal{K}=\left[\ket{K_{0}},\ket{K_{1}},\ket{K_{2}},\ket{K_{3}},\ket{K_{4}},\ket{K_{5}},\ket{K_{6}},\ket{K_{7}},\ket{K_{8}}\right],
\label{eq:collisionmatrix1}
\end{equation}
which can be explicitly written as
\begin{equation}
\mathcal{K}= \left[
\begin{array}{rrrrrrrrr}
1 & 0  &  0 & -4 & 0  & 0  & 0 & 0 &  4\\
1 & 1  &  0 & -1 & 1  & 0  & 0 & 2 & -2\\
1 & 0  &  1 & -1 & -1 & 0  & 2 & 0 & -2\\
1 & -1 &  0 & -1 & 1  & 0  & 0 &-2 & -2\\
1 & 0  & -1 & -1 & -1 & 0  &-2 & 0 & -2\\
1 & 1  &  1 &  2 & 0  & 1  &-1 &-1 &  1\\
1 & -1 &  1 &  2 & 0  & -1 &-1 & 1 &  1\\
1 & -1 & -1 &  2 & 0  & 1  & 1 & 1 &  1\\
1 & 1  & -1 &  2 & 0  & -1 & 1 &-1 &  1\\
\end{array} \right].
\label{eq:collisionmatrix2}
\end{equation}
It possesses a number of interesting properties including a computationally useful fact that
$\mathcal{K}\mathcal{K}^\dag$ is a diagonal matrix.

\section{\label{sec:ccentraleqmmoments}Continuous Central Moments: Equilibrium and Sources}
Consider an athermal fluid in motion which is characterized by its local hydrodynamic fields at the Cartesian coordinate $(x,y)$, i.e.
density $\rho$, hydrodynamic velocity $\overrightarrow{u}=(u_x,u_y)$, and subjected to a force field $\overrightarrow{F}=(F_x,F_y)$, whose origin could be either internal or external to the system. The local Maxwell-Boltzmann distribution, or, simply, the Maxwellian in \emph{continuous} particle velocity space $(\xi_x,\xi_y)$ is given by
\begin{equation}
f^\mathcal{M}\equiv
f^\mathcal{M}(\rho,\overrightarrow{u},\xi_x,\xi_y)=\frac{\rho}{2\pi c_s^2}\exp\left[-\frac{\left(\overrightarrow{\xi}-\overrightarrow{u}\right)^2}{2c_s^2}\right],\label{eq:Maxwellian}
\end{equation}
where we choose
\begin{equation}
c_s^2=1/3.
\end{equation}

Let us now define \emph{continuous} central moments, i.e. moments displaced by the local hydrodynamic velocity, of order $(m+n)$:
\begin{equation}
\widehat{\Pi}^{\mathcal{M}}_{x^my^n}=\int_{-\infty}^{\infty}\int_{-\infty}^{\infty}f^\mathcal{M}(\xi_x-u_x)^m(\xi_y-u_y)^nd\xi_xd\xi_y.
\end{equation}
By virtue of the fact that $f^\mathcal{M}$ being an even function, $\widehat{\Pi}^{\mathcal{M}}_{x^my^n}\neq0$ when $m$ and $n$ are even and $\widehat{\Pi}^{\mathcal{M}}_{x^my^n}=0$ when $m$ or $n$ odd. Here and henceforth, the subscripts $x^my^n$ mean $xxx\cdots m-\text{times}$ and $yyy\cdots n-\text{times}$. Thus, evaluating this quantity in the increasing order of moments gives
\begin{eqnarray*}
\widehat{\Pi}^{\mathcal{M}}_{0}&=&\rho, \\
\widehat{\Pi}^{\mathcal{M}}_{x}&=&0, \\
\widehat{\Pi}^{\mathcal{M}}_{y}&=&0, \\
\widehat{\Pi}^{\mathcal{M}}_{xx}&=&c_s^2\rho,\\
\widehat{\Pi}^{\mathcal{M}}_{yy}&=&c_s^2\rho,\\
\widehat{\Pi}^{\mathcal{M}}_{xy}&=&0, \\
\widehat{\Pi}^{\mathcal{M}}_{xxy}&=&0, \\
\widehat{\Pi}^{\mathcal{M}}_{xyy}&=&0, \\
\widehat{\Pi}^{\mathcal{M}}_{xxyy}&=&c_s^4\rho.
\end{eqnarray*}
Here, and in the rest of this paper, the use of ``hat" over a symbol represents values in the space of moments.

Now, we propose that the continuous distribution function $f$ is modified by the presence of a force field as given by the following ansatz:
\begin{equation}
\Delta f^{F}=\frac{\overrightarrow{F}}{\rho}\cdot\frac{(\overrightarrow{\xi}-\overrightarrow{u})}{c_s^2}f^\mathcal{M}
\label{eq:forceansatz}
\end{equation}
It may be noted that He \emph{et al}. (1998)~\cite{he98} proposed similar form for the continuous Boltzmann equation to derive source terms for the SRT-LBE. However, it's influence on discrete distribution function due to cascaded collision process via the method of central moments to establish Galilean invariant solutions is expected to be, in general, be different. Let us now define a corresponding \emph{continuous} central moment of order $(m+n)$ due to change in the distribution function as a result of a force field as
\begin{equation}
\widehat{\Gamma}^{F}_{x^my^n}=\int_{-\infty}^{\infty}\int_{-\infty}^{\infty}\Delta f^F(\xi_x-u_x)^m(\xi_y-u_y)^nd\xi_xd\xi_y.
\label{eq:centralmomentforce}
\end{equation}
Evaluation of Eq.~(\ref{eq:centralmomentforce}) in the increasing order of moments yields
\begin{eqnarray*}
\widehat{\Gamma}^{F}_{0}&=&0, \\
\widehat{\Gamma}^{F}_{x}&=&F_x, \\
\widehat{\Gamma}^{F}_{y}&=&F_y,\\
\widehat{\Gamma}^{F}_{xx}&=&0,\\
\widehat{\Gamma}^{F}_{yy}&=&0,\\
\widehat{\Gamma}^{F}_{xy}&=&0,\\
\widehat{\Gamma}^{F}_{xxy}&=&c_s^2F_y,\\
\widehat{\Gamma}^{F}_{xyy}&=&c_s^2F_x,\\
\widehat{\Gamma}^{F}_{xxyy}&=&0.
\end{eqnarray*}

\section{\label{sec:cascadedLBEforcing}Cascaded Lattice-Boltzmann Method with Forcing Terms}
First, let us define a \emph{discrete} distribution function supported by the discrete particle velocity set $\overrightarrow{e}_{\alpha}$:
\begin{equation}
\mathbf{f}=\ket{f_{\alpha}}=(f_0,f_1,f_2,\ldots,f_8)^\dag.
\end{equation}
Following Geier \emph{et al}.~\cite{geier06}, we represent collision as a cascaded process in which the effect of collision on lower order moments successively influences those of higher order in a cascaded manner. In particular, we model the change in discrete distribution due to collision as
\begin{equation}
\Omega_{\alpha}^{c}\equiv \Omega_{\alpha}^{c}(\mathbf{f},\mathbf{\widehat{g}})=(\mathcal{K}\cdot \mathbf{\widehat{g}})_{\alpha},
\label{eq:cascadedcollisionoperator}
\end{equation}
where
\begin{equation}
\mathbf{\widehat{g}}=\ket{\widehat{g}_{\alpha}}=(\widehat{g}_0,\widehat{g}_1,\widehat{g}_2,\ldots,\widehat{g}_8)^\dag
\end{equation}
determines the changes in \emph{discrete} moment space in a cascaded manner. That is, in general,
\begin{equation}
\widehat{g}_{\alpha}\equiv\widehat{g}_{\alpha}(\mathbf{f},\widehat{g}_{\beta}), \qquad \beta=0,1,2,\ldots,\alpha-1.
\end{equation}
The detailed structure of $\mathbf{\widehat{g}}$ will be determined later in Sec.~\ref{sec:cascadedcollisionforcing}.

We define that $f_{\alpha}$ changes due to external force field $\overrightarrow{F}$ by the \emph{discrete} source term $S_{\alpha}$. That is,
\begin{equation}
\mathbf{S}=\ket{S_{\alpha}}=(S_0,S_1,S_2,\ldots,S_8)^\dag.
\end{equation}
We suppose that particle populations are continuously affected by this in time as they traverse along their characteristics. The precise form of $S_{\alpha}$ is yet unknown and will be determined as part of the procedure presented in Sec.~\ref{sec:cascadedcollisionforcing}.

With the above definitions, the evolution of $f_{\alpha}$ in the Cascaded-LBM can be written as
\begin{equation}
f_{\alpha}(\overrightarrow{x}+\overrightarrow{e}_{\alpha},t+1)=f_{\alpha}(\overrightarrow{x},t)+\Omega_{{\alpha}(\overrightarrow{x},t)}^{c}+
\int_{t}^{t+1}S_{{\alpha}(\overrightarrow{x}+\overrightarrow{e}_{\alpha}\theta,t+\theta)}d\theta,
\label{eq:cascadedLBE1}
\end{equation}
where the fluid dynamical variables are determined by
\begin{eqnarray}
\rho&=&\sum_{\alpha=0}^{8}f_{\alpha}=\braket{f_{\alpha}|\rho},\\
\rho u_i&=&\sum_{\alpha=0}^{8}f_{\alpha} e_{\alpha i}=\braket{f_{\alpha}|e_{\alpha i}}, i \in {x,y}.
\end{eqnarray}

The last term on the right-hand-side (RHS) of Eq.~(\ref{eq:cascadedLBE1}) represents the cumulative effect of forces as particle populations advect along their characteristic directions. Various approaches are possible here to numerically represent this integral, with the simplest being an explicit rule. However, in general cases where $\overrightarrow{F}$ can have spatial and temporal dependencies, for improved accuracy, it becomes imperative to represent it with a higher order scheme. One common approach, which is employed here, is to apply a second-order trapezoidal rule, which will sample both the temporal end points, $(t,t+1)$, along the characteristic direction $\alpha$. That is,
\begin{equation}
f_{\alpha}(\overrightarrow{x}+\overrightarrow{e}_{\alpha},t+1)=f_{\alpha}(\overrightarrow{x},t)+\Omega_{{\alpha}(\overrightarrow{x},t)}^{c}+
\frac{1}{2}\left[S_{{\alpha}(\overrightarrow{x},t)}+S_{{\alpha}(\overrightarrow{x}+\overrightarrow{e}_{\alpha},t+1)}\right]
\label{eq:cascadedLBE2}
\end{equation}

Equation (\ref{eq:cascadedLBE2}) is semi-implicit. To remove implicitness along discrete characteristics, we apply the following transformation~\cite{he98,premnath06}:
\begin{equation}
\overline{f}_{\alpha}=f_{\alpha}-\frac{1}{2}S_{\alpha}.
\label{eq:transformation}
\end{equation}
Thus, Eq.~(\ref{eq:cascadedLBE2}) becomes
\begin{equation}
\overline{f}_{\alpha}(\overrightarrow{x}+\overrightarrow{e}_{\alpha},t+1)=\overline{f}_{\alpha}(\overrightarrow{x},t)+\Omega_{{\alpha}(\overrightarrow{x},t)}^{c}+
S_{{\alpha}(\overrightarrow{x},t)}.
\label{eq:cascadedLBE3}
\end{equation}
Clearly, we need to determine $\sum_{\alpha}S_{\alpha}$ and $\sum_{\alpha}S_{\alpha}\overrightarrow{e}_{\alpha}$ to obtain $\rho$ and $\rho \overrightarrow{u}$, respectively, in terms of the transformed variable $\overline{f}_{\alpha}$, which will be carried out in the next section.

\section{\label{sec:cascadedcollisionforcing}Construction of Cascaded Collision Operator and Forcing Terms}
In order to determine the structure of the cascaded collision operator and the source terms in the presence of force fields, we now define the following \emph{discrete} central moments of the distribution functions and source terms, respectively:
\begin{eqnarray}
\widehat{\kappa}_{x^m y^n}&=&\sum_{\alpha}f_{\alpha}(e_{\alpha x}-u_x)^m(e_{\alpha y}-u_y)^n=\braket{(e_{\alpha x}-u_x)^m(e_{\alpha y}-u_y)^n|f_{\alpha}},\label{eq:centralmomentdistributionfunction1}\\
\widehat{\sigma}_{x^m y^n}&=&\sum_{\alpha}S_{\alpha}(e_{\alpha x}-u_x)^m(e_{\alpha y}-u_y)^n=\braket{(e_{\alpha x}-u_x)^m(e_{\alpha y}-u_y)^n|S_{\alpha}}\label{eq:centralmomentforcingterm1}.
\end{eqnarray}
We also define a \emph{discrete} central moment in terms of transformed distribution function to facilitate subsequent calculations:
\begin{equation}
\widehat{\overline{\kappa}}_{x^m y^n}=\sum_{\alpha}\overline{f}_{\alpha}(e_{\alpha x}-u_x)^m(e_{\alpha y}-u_y)^n=\braket{(e_{\alpha x}-u_x)^m(e_{\alpha y}-u_y)^n|\overline{f}_{\alpha}}.
\end{equation}
Owing to Eq.~(\ref{eq:transformation}), it follows that
\begin{equation}
\widehat{\overline{\kappa}}_{x^m y^n}=\widehat{\kappa}_{x^m y^n}-\frac{1}{2}\widehat{\sigma}_{x^m y^n}.
\label{eq:transformedcentralmoments}
\end{equation}

Let us also suppose that $f_{\alpha}$ and $\overline{f}_{\alpha}$ have certain local equilibrium states represented by $f_{\alpha}^{eq}$ and $\overline{f}_{\alpha}^{eq}$, respectively, and the corresponding central moments are
\begin{eqnarray}
\widehat{\kappa}_{x^m y^n}^{eq}&=&\sum_{\alpha}f_{\alpha}^{eq}(e_{\alpha x}-u_x)^m(e_{\alpha y}-u_y)^n=\braket{(e_{\alpha x}-u_x)^m(e_{\alpha y}-u_y)^n|f_{\alpha}^{eq}},\\
\widehat{\overline{\kappa}}_{x^m y^n}^{eq}&=&\sum_{\alpha}\overline{f}_{\alpha}^{eq}(e_{\alpha x}-u_x)^m(e_{\alpha y}-u_y)^n=\braket{(e_{\alpha x}-u_x)^m(e_{\alpha y}-u_y)^n|\overline{f}_{\alpha}^{eq}}.
\end{eqnarray}

Now, we take an important step by equating the \emph{discrete} central moments for both the distribution functions (equilibrium) and source terms, defined above, with the \emph{continuous} central moments derived in Sec.~\ref{sec:ccentraleqmmoments}. Thus, we have
\begin{eqnarray}
\widehat{\kappa}_{x^m y^n}^{eq}&=&\widehat{\Pi}^{\mathcal{M}}_{x^m y^n},\\
\widehat{\sigma}_{x^m y^n}&=&\widehat{\Gamma}^{F}_{x^m y^n}.
\end{eqnarray}
In other words, the discrete central moments of various orders for both the distribution functions (equilibrium) and source terms, respectively, become
\begin{eqnarray}
\widehat{\kappa}^{eq}_{0}&=&\rho, \\
\widehat{\kappa}^{eq}_{x}&=&0, \\
\widehat{\kappa}^{eq}_{y}&=&0, \\
\widehat{\kappa}^{eq}_{xx}&=&c_s^2\rho,\\
\widehat{\kappa}^{eq}_{yy}&=&c_s^2\rho,\\
\widehat{\kappa}^{eq}_{xy}&=&0, \\
\widehat{\kappa}^{eq}_{xxy}&=&0, \\
\widehat{\kappa}^{eq}_{xyy}&=&0, \\
\widehat{\kappa}^{eq}_{xxyy}&=&c_s^4\rho,
\end{eqnarray}
and
\begin{eqnarray}
\widehat{\sigma}_{0}&=&0, \label{eq:centralmomentforcing0}\\
\widehat{\sigma}_{x}&=&F_x, \label{eq:centralmomentforcing1}\\
\widehat{\sigma}_{y}&=&F_y,\label{eq:centralmomentforcing2}\\
\widehat{\sigma}_{xx}&=&0, \label{eq:centralmomentforcing3}\\
\widehat{\sigma}_{yy}&=&0, \label{eq:centralmomentforcing4}\\
\widehat{\sigma}_{xy}&=&0, \label{eq:centralmomentforcing5}\\
\widehat{\sigma}_{xxy}&=&c_s^2F_y, \label{eq:centralmomentforcing6}\\
\widehat{\sigma}_{xyy}&=&c_s^2F_x, \label{eq:centralmomentforcing7}\\
\widehat{\sigma}_{xxyy}&=&0.\label{eq:centralmomentforcing8}
\end{eqnarray}
From Eq.~(\ref{eq:transformedcentralmoments}), we get the following transformed central moments, which comprises as one of the main elements for subsequent development and analysis:
\begin{eqnarray}
\widehat{\overline{\kappa}}^{eq}_{0}&=&\rho, \label{eq:centralmomenttransformed0}\\
\widehat{\overline{\kappa}}^{eq}_{x}&=&-\frac{1}{2}F_x, \label{eq:centralmomenttransformed1}\\
\widehat{\overline{\kappa}}^{eq}_{y}&=&-\frac{1}{2}F_y, \label{eq:centralmomenttransformed2}\\
\widehat{\overline{\kappa}}^{eq}_{xx}&=&c_s^2\rho,\label{eq:centralmomenttransformed3}\\
\widehat{\overline{\kappa}}^{eq}_{yy}&=&c_s^2\rho,\label{eq:centralmomenttransformed4}\\
\widehat{\overline{\kappa}}^{eq}_{xy}&=&0, \label{eq:centralmomenttransformed5}\\
\widehat{\overline{\kappa}}^{eq}_{xxy}&=&-\frac{c_s^2}{2}F_y, \label{eq:centralmomenttransformed6}\\
\widehat{\overline{\kappa}}^{eq}_{xyy}&=&-\frac{c_s^2}{2}F_x, \label{eq:centralmomenttransformed7}\\
\widehat{\overline{\kappa}}^{eq}_{xxyy}&=&c_s^4\rho. \label{eq:centralmomenttransformed8}
\end{eqnarray}

To proceed further, we need to obtain the corresponding moments in rest or lattice frame of reference, i.e. raw moments. The tool that we employ for this purpose is the binomial theorem. The transformation between the central moments and the raw moments for any state variable $\varphi$ supported by discrete particle velocity set can be formally written as
\begin{eqnarray}
\braket{(e_{\alpha x}-u_x)^m(e_{\alpha y}-u_y)^n|\varphi}&=&\braket{e_{\alpha x}^m e_{\alpha y}^n|\varphi}+
\braket{e_{\alpha x}^m \left[\sum_{j=1}^{n} C^n_je_{\alpha y}^{n-j}(-1)^{j}u_y^{j}\right]|\varphi}\nonumber +\\
&& \braket{e_{\alpha y}^n \left[\sum_{i=1}^{m} C^m_ie_{\alpha x}^{m-i}(-1)^{i}u_x^{i}\right]|\varphi}\nonumber +\\
&& \braket{ \left[\sum_{i=1}^{m} C^m_ie_{\alpha x}^{m-i}(-1)^{i}u_x^{i}\right]\left[\sum_{j=1}^{n} C^n_je_{\alpha y}^{n-j}(-1)^{j}u_y^{j}\right]|\varphi}
\label{eq:binomialtheorem}
\end{eqnarray}
where
$C^{p}_{q}=p!/(q!(p-q)!)$. In the above, commutation of the inner product of vectors, represented using the ``bra-ket" operators, with summations and scalar products is assumed. Clearly, raw moments of equal or lesser order in combination is equivalent to central moments of a given order.

Application of Eq.~(\ref{eq:binomialtheorem}) to the forcing terms, i.e., using Eq.~(\ref{eq:centralmomentforcingterm1}) and
Eqs.~(\ref{eq:centralmomentforcing0})-(\ref{eq:centralmomentforcing8}) yields analytical expressions in the rest frame of reference:
\begin{eqnarray}
\braket{S_{\alpha}|\rho}&=&\sum_{\alpha}S_{\alpha}=0,\label{eq:rawmomentsourceterm0}\\
\braket{S_{\alpha}|e_{\alpha x}}&=&\sum_{\alpha}S_{\alpha} e_{\alpha x}=F_x,\label{eq:rawmomentsourceterm1}\\
\braket{S_{\alpha}|e_{\alpha y}}&=&\sum_{\alpha}S_{\alpha} e_{\alpha y}=F_y,\label{eq:rawmomentsourceterm2}\\
\braket{S_{\alpha}|e_{\alpha x}^2}&=&\sum_{\alpha}S_{\alpha} e_{\alpha x}^2=2F_xu_x,\label{eq:rawmomentsourceterm3}\\
\braket{S_{\alpha}|e_{\alpha y}^2}&=&\sum_{\alpha}S_{\alpha} e_{\alpha y}^2=2F_xu_y,\label{eq:rawmomentsourceterm4}\\
\braket{S_{\alpha}|e_{\alpha x}e_{\alpha y}}&=&\sum_{\alpha}S_{\alpha} e_{\alpha x}e_{\alpha y}=F_xu_y+F_yu_x,\label{eq:rawmomentsourceterm5}\\
\braket{S_{\alpha}|e_{\alpha x}^2e_{\alpha y}}&=&\sum_{\alpha}S_{\alpha} e_{\alpha x}^2e_{\alpha y}=\left(\frac{1}{3}+u_x^2\right)F_y+2F_xu_xu_y,\label{eq:rawmomentsourceterm6}\\
\braket{S_{\alpha}|e_{\alpha x}e_{\alpha y}^2}&=&\sum_{\alpha}S_{\alpha} e_{\alpha x}e_{\alpha y}^2=\left(\frac{1}{3}+u_y^2\right)F_x+2F_yu_yu_x,\label{eq:rawmomentsourceterm7}\\
\braket{S_{\alpha}|e_{\alpha x}^2e_{\alpha y}^2}&=&\sum_{\alpha}S_{\alpha} e_{\alpha x}^2e_{\alpha y}^2=\left(\frac{2}{3}+2u_y^2\right)F_xu_x+\left(\frac{2}{3}+2u_x^2\right)F_yu_y.\label{eq:rawmomentsourceterm8}
\end{eqnarray}

For subsequent procedure, we also need the raw moments of the collision kernel
\begin{equation}
\sum_{\alpha}(\mathcal{K}\cdot \mathbf{\widehat{g}})_{\alpha}e_{\alpha x}^m e_{\alpha y}^n = \sum_{\beta} \braket{K_{\beta}|e_{\alpha x}^m e_{\alpha y}^n}\widehat{g}_{\beta}.
\end{equation}
Since collisions do not change mass and momenta, which are thus called collisional invariants, we can set
\begin{equation}
\widehat{g}_{0}=\widehat{g}_{1}=\widehat{g}_{2}.
\end{equation}
Thus, we effectively need to determine the functional expressions for $\widehat{g}_{\beta}$ for $\beta=3,4,\ldots,8$. Owing to the \emph{orthogonal} property of the eigenvectors of $\mathcal{K}$ by construction,  i.e. Eq.~(\ref{eq:collisionmatrix1}), we obtain
\begin{eqnarray}
\sum_{\alpha}(\mathcal{K}\cdot \mathbf{\widehat{g}})_{\alpha}=
\sum_{\beta} \braket{K_{\beta}|\rho}\widehat{g}_{\beta}&=&0,\label{eq:collisionkernelmoment0}\\
\sum_{\alpha}(\mathcal{K}\cdot \mathbf{\widehat{g}})_{\alpha}e_{\alpha x}=
\sum_{\beta} \braket{K_{\beta}|e_{\alpha x}}\widehat{g}_{\beta}&=&0,\label{eq:collisionkernelmoment1}\\
\sum_{\alpha}(\mathcal{K}\cdot \mathbf{\widehat{g}})_{\alpha}e_{\alpha y}=
\sum_{\beta} \braket{K_{\beta}|e_{\alpha y}}\widehat{g}_{\beta}&=&0,\label{eq:collisionkernelmoment2}\\
\sum_{\alpha}(\mathcal{K}\cdot \mathbf{\widehat{g}})_{\alpha}e_{\alpha x}^2=
\sum_{\beta} \braket{K_{\beta}|e_{\alpha x}^2}\widehat{g}_{\beta}&=&6\widehat{g}_{3}+2\widehat{g}_{4},\label{eq:collisionkernelmoment3}\\
\sum_{\alpha}(\mathcal{K}\cdot \mathbf{\widehat{g}})_{\alpha}e_{\alpha y}^2=
\sum_{\beta} \braket{K_{\beta}|e_{\alpha y}^2}\widehat{g}_{\beta}&=&6\widehat{g}_{3}-2\widehat{g}_{4},\label{eq:collisionkernelmoment4}\\
\sum_{\alpha}(\mathcal{K}\cdot \mathbf{\widehat{g}})_{\alpha}e_{\alpha x}e_{\alpha y}=
\sum_{\beta} \braket{K_{\beta}|e_{\alpha x}e_{\alpha y}}\widehat{g}_{\beta}&=&4\widehat{g}_{5},\label{eq:collisionkernelmoment5}\\
\sum_{\alpha}(\mathcal{K}\cdot \mathbf{\widehat{g}})_{\alpha}e_{\alpha x}^2e_{\alpha y}=
\sum_{\beta} \braket{K_{\beta}|e_{\alpha x}^2e_{\alpha y}}\widehat{g}_{\beta}&=&-4\widehat{g}_{6},\label{eq:collisionkernelmoment6}\\
\sum_{\alpha}(\mathcal{K}\cdot \mathbf{\widehat{g}})_{\alpha}e_{\alpha x}e_{\alpha y}^2=
\sum_{\beta} \braket{K_{\beta}|e_{\alpha x}e_{\alpha y}^2}\widehat{g}_{\beta}&=&-4\widehat{g}_{7},\label{eq:collisionkernelmoment7}\\
\sum_{\alpha}(\mathcal{K}\cdot \mathbf{\widehat{g}})_{\alpha}e_{\alpha x}^2e_{\alpha y}^2=
\sum_{\beta} \braket{K_{\beta}|e_{\alpha x}^2e_{\alpha y}^2}\widehat{g}_{\beta}&=&8\widehat{g}_{3}+4\widehat{g}_{8}. \label{eq:collisionkernelmoment8}
\end{eqnarray}

Now, for computational convenience, the evolution equation, Eq.~(\ref{eq:cascadedLBE3}), of the Cascaded-LBM with forcing term may be rewritten as
\begin{eqnarray}
\widetilde{\overline{f}}_{\alpha}(\overrightarrow{x},t)&=&\overline{f}_{\alpha}(\overrightarrow{x},t)+\Omega_{{\alpha}(\overrightarrow{x},t)}^{c}+
S_{{\alpha}(\overrightarrow{x},t)},\label{eq:cascadedcollision1}\\
\overline{f}_{\alpha}(\overrightarrow{x}+\overrightarrow{e}_{\alpha},t+1)&=&\widetilde{\overline{f}}_{\alpha}(\overrightarrow{x},t). \label{eq:cascadedstreaming1}
\end{eqnarray}
where Eq.~(\ref{eq:cascadedcollision1}) and Eq.~(\ref{eq:cascadedstreaming1}) represent the collision step, augmented by forcing term, and streaming step, respectively. Here and henceforth, the symbol ``tilde" ($\sim$) refers to the post-collision state. The hydrodynamic variables can then be obtained as
\begin{eqnarray}
\rho&=&\sum_{\alpha=0}^{8}\overline{f}_{\alpha}=\braket{\overline{f}_{\alpha}|\rho},\label{eq:densitycalculation}\\
\rho u_i&=&\sum_{\alpha=0}^{8}\overline{f}_{\alpha} e_{\alpha i}+\frac{1}{2}F_i=\braket{\overline{f}_{\alpha}|e_{\alpha i}}+\frac{1}{2}F_i, i \in {x,y} \label{eq:velocitycalculation}
\end{eqnarray}
in view of Eqs.~(\ref{eq:transformation}),~(\ref{eq:rawmomentsourceterm0}),~(\ref{eq:rawmomentsourceterm1})~and~(\ref{eq:rawmomentsourceterm2}).

Now, to obtain the source terms in particle velocity space, we first compute $\braket{K_{\beta}|S_{\alpha}}$, $\beta=0,1,2,\ldots,8$. From Eqs.~(\ref{eq:collisionmatrix1}) and (\ref{eq:rawmomentsourceterm0})-(\ref{eq:rawmomentsourceterm8}), we readily get
\begin{eqnarray}
\widehat{m}^{s}_{0}=\braket{K_0|S_{\alpha}}&=&0, \label{eq:sourceterm0}\\
\widehat{m}^{s}_{1}=\braket{K_1|S_{\alpha}}&=&F_x, \label{eq:sourceterm1}\\
\widehat{m}^{s}_{2}=\braket{K_2|S_{\alpha}}&=&F_y, \label{eq:sourceterm2}\\
\widehat{m}^{s}_{3}=\braket{K_3|S_{\alpha}}&=&6(F_xu_x+F_yu_y), \label{eq:sourceterm3}\\
\widehat{m}^{s}_{4}=\braket{K_4|S_{\alpha}}&=&2(F_xu_x-F_yu_y), \label{eq:sourceterm4}\\
\widehat{m}^{s}_{5}=\braket{K_5|S_{\alpha}}&=&(F_xu_y+F_yu_x), \label{eq:sourceterm5}\\
\widehat{m}^{s}_{6}=\braket{K_6|S_{\alpha}}&=&(1-3u_x^2)F_y-6F_xu_xu_y, \label{eq:sourceterm6}\\
\widehat{m}^{s}_{7}=\braket{K_7|S_{\alpha}}&=&(1-3u_y^2)F_x-6F_yu_yu_x, \label{eq:sourceterm7}\\
\widehat{m}^{s}_{8}=\braket{K_8|S_{\alpha}}&=&3\left[(6u_y^2-2)F_xu_x+(6u_x^2-2)F_yu_y\right]. \label{eq:sourceterm8}
\end{eqnarray}

Thus, we can write
\begin{eqnarray}
(\mathcal{K}\cdot\mathbf{S})_{\alpha}&=&(\braket{K_0|S_{\alpha}},\braket{K_1|S_{\alpha}},\braket{K_2|S_{\alpha}},\ldots,\braket{K_8|S_{\alpha}}) \nonumber \\
&=& (\widehat{m}^{s}_{0},\widehat{m}^{s}_{1},\widehat{m}^{s}_{2},\ldots,\widehat{m}^{s}_{8})^T\equiv \ket{\widehat{m}^{s}_{\alpha}}. \label{eq:sourceformulation1}
\end{eqnarray}
By virtue of orthogonality of $\mathcal{K}$, we have
$ \mathcal{K}\mathcal{K}^\dag=~D~\equiv\text{diag}(\braket{K_0|K_0},\braket{K_1|K_1},\braket{K_2|K_2},\ldots,\braket{K_8|K_8})=\text{diag}(9,6,6,36,4,4,12,12,36)$. Inverting Eq.~(\ref{eq:sourceformulation1}) by making use of the property $\mathcal{K}^{-1}=\mathcal{K}^\dag \cdot D^{-1}$, we get explicit expressions for $S_{\alpha}$ in terms of $\overrightarrow{F}$ and $\overrightarrow{u}$ in particle velocity space as
\begin{eqnarray}
S_0&=&\frac{1}{9}\left[-\widehat{m}^{s}_{3}+\widehat{m}^{s}_{8}\right], \label{eq:vsourceterms0}\\
S_1&=&\frac{1}{36}\left[6\widehat{m}^{s}_{1}-\widehat{m}^{s}_{3}+9\widehat{m}^{s}_{4}+6\widehat{m}^{s}_{7}-2\widehat{m}^{s}_{8}\right], \label{eq:vsourceterms1}\\
S_2&=&\frac{1}{36}\left[6\widehat{m}^{s}_{2}-\widehat{m}^{s}_{3}-9\widehat{m}^{s}_{4}+6\widehat{m}^{s}_{6}-2\widehat{m}^{s}_{8}\right], \label{eq:vsourceterms2}\\
S_3&=&\frac{1}{36}\left[-6\widehat{m}^{s}_{1}-\widehat{m}^{s}_{3}+9\widehat{m}^{s}_{4}-6\widehat{m}^{s}_{7}-2\widehat{m}^{s}_{8}\right], \label{eq:vsourceterms3}\\
S_4&=&\frac{1}{36}\left[-6\widehat{m}^{s}_{2}-\widehat{m}^{s}_{3}-9\widehat{m}^{s}_{4}-6\widehat{m}^{s}_{6}-2\widehat{m}^{s}_{8}\right], \label{eq:vsourceterms4}\\
S_5&=&\frac{1}{36}\left[6\widehat{m}^{s}_{1}+6\widehat{m}^{s}_{2}+2\widehat{m}^{s}_{3}
+9\widehat{m}^{s}_{5}-3\widehat{m}^{s}_{6}-3\widehat{m}^{s}_{7}+\widehat{m}^{s}_{8}\right], \label{eq:vsourceterms5}\\
S_6&=&\frac{1}{36}\left[-6\widehat{m}^{s}_{1}+6\widehat{m}^{s}_{2}+2\widehat{m}^{s}_{3}
-9\widehat{m}^{s}_{5}-3\widehat{m}^{s}_{6}+3\widehat{m}^{s}_{7}+\widehat{m}^{s}_{8}\right], \label{eq:vsourceterms6}\\
S_7&=&\frac{1}{36}\left[-6\widehat{m}^{s}_{1}-6\widehat{m}^{s}_{2}+2\widehat{m}^{s}_{3}
+9\widehat{m}^{s}_{5}+3\widehat{m}^{s}_{6}+3\widehat{m}^{s}_{7}+\widehat{m}^{s}_{8}\right], \label{eq:vsourceterms7}\\
S_8&=&\frac{1}{36}\left[6\widehat{m}^{s}_{1}-6\widehat{m}^{s}_{2}+2\widehat{m}^{s}_{3}
-9\widehat{m}^{s}_{5}+3\widehat{m}^{s}_{6}-3\widehat{m}^{s}_{7}+\widehat{m}^{s}_{8}\right]. \label{eq:vsourceterms8}
\end{eqnarray}

We now need to find the expressions of $\braket{\overline{f}_{\alpha}|e_{\alpha x}^m e_{\alpha y}^n}=\sum_{\alpha=0}^{8}\overline{f}_{\alpha}e_{\alpha x}^m e_{\alpha y}^n$ to proceed further. In this regard,
for convenience, we define the following notation for a compact summation operator acting on the transformed distribution function $\overline{f}_{\alpha}$:
\begin{equation}
a(\overline{f}_{\alpha_1}+\overline{f}_{\alpha_3}+\overline{f}_{\alpha_3}+\cdots)+
b(\overline{f}_{\beta_1}+\overline{f}_{\beta_2}+\overline{f}_{\beta_3}+\cdots)+\cdots=\left(a\sum_{\alpha}^A+b\sum_{\alpha}^B+\cdots\right)\otimes \overline{f}_{\alpha},
\end{equation}
where $A=\left\{\alpha_1,\alpha_2,\alpha_3,\cdots\right\}$,~$B=\left\{\beta_1,\beta_2,\beta_3,\cdots\right\}$,$\cdots$. For conserved basis vectors, we have them in terms of collisional invariants
\begin{eqnarray}
\braket{\overline{f}_{\alpha}|\rho}=\sum_{\alpha = 0}^{8}\overline{f}_{\alpha}&=&\rho,\\
\braket{\overline{f}_{\alpha}|e_{\alpha x}}=\sum_{\alpha = 0}^{8}\overline{f}_{\alpha} e_{\alpha x}&=&\rho u_x-\frac{1}{2}F_x,\\
\braket{\overline{f}_{\alpha}|e_{\alpha y}}=\sum_{\alpha = 0}^{8}\overline{f}_{\alpha} e_{\alpha y}&=&\rho u_y-\frac{1}{2}F_y,
\end{eqnarray}
and, for the non-conserved basis vectors, we have
\begin{eqnarray}
\braket{\overline{f}_{\alpha}|e_{\alpha x}^2}=\sum_{\alpha = 0}^{8}\overline{f}_{\alpha} e_{\alpha x}^2&=&\left(\sum_{\alpha}^{A_3}\right)\otimes \overline{f}_{\alpha},\label{eq:nonconservedbasisvector3}\\
\braket{\overline{f}_{\alpha}|e_{\alpha y}^2}=\sum_{\alpha = 0}^{8}\overline{f}_{\alpha} e_{\alpha y}^2&=&\left(\sum_{\alpha}^{A_4}\right)\otimes \overline{f}_{\alpha},\label{eq:nonconservedbasisvector4}\\
\braket{\overline{f}_{\alpha}|e_{\alpha x}e_{\alpha y}}=\sum_{\alpha = 0}^{8}\overline{f}_{\alpha} e_{\alpha x} e_{\alpha y}&=&\left(\sum_{\alpha}^{A_5}-\sum_{\alpha}^{B_5}\right)\otimes \overline{f}_{\alpha},\label{eq:nonconservedbasisvector5}\\
\braket{\overline{f}_{\alpha}|e_{\alpha x}^2e_{\alpha y}}=\sum_{\alpha = 0}^{8}\overline{f}_{\alpha} e_{\alpha x}^2 e_{\alpha y}&=&\left(\sum_{\alpha}^{A_6}-\sum_{\alpha}^{B_6}\right)\otimes \overline{f}_{\alpha},\label{eq:nonconservedbasisvector6}\\
\braket{\overline{f}_{\alpha}|e_{\alpha x}e_{\alpha y}^2}=\sum_{\alpha = 0}^{8}\overline{f}_{\alpha} e_{\alpha x} e_{\alpha y}^2&=&\left(\sum_{\alpha}^{A_7}-\sum_{\alpha}^{B_7}\right)\otimes \overline{f}_{\alpha},\label{eq:nonconservedbasisvector7}\\
\braket{\overline{f}_{\alpha}|e_{\alpha x}^2e_{\alpha y}^2}=\sum_{\alpha = 0}^{8}\overline{f}_{\alpha} e_{\alpha x}^2 e_{\alpha y}^2&=&\left(\sum_{\alpha}^{A_8}\right)\otimes \overline{f}_{\alpha},\label{eq:nonconservedbasisvector8}
\end{eqnarray}
where
\begin{eqnarray}
A_3&=&\left\{1,3,5,6,7,8\right\},\\
A_4&=&\left\{2,4,5,6,7,8\right\},\\
A_5&=&\left\{5,7\right\},B_5=\left\{6,8\right\},\\
A_6&=&\left\{5,6\right\},B_6=\left\{7,8\right\},\\
A_7&=&\left\{5,8\right\},B_7=\left\{6,7\right\},\\
A_8&=&\left\{5,6,7,8\right\}.
\end{eqnarray}

With the above preliminaries, we are now in a position to determine the structure of the cascaded collision operator in the presence of forcing terms. Starting from the lowest order non-conservative post-collision central moments, we successively set them equal to their corresponding equilibrium states. Once the expressions for $\widehat{g}_{\beta}$ is determined, we discard this equilibrium assumption and multiply it with a corresponding relaxation parameter to allow for a relaxation process during collision~\cite{geier06}. From Eq.~(\ref{eq:centralmomenttransformed3}), which is the lowest non-conserved central moment, and applying the binomial theorem (Eq.~(\ref{eq:binomialtheorem})) to transform it to the rest frame of reference, we get
\begin{equation}
\widehat{\overline{\kappa}}_{xx}^{eq}=1/3\rho=\braket{\widetilde{\overline{f}}_{\alpha}|e_{\alpha x}^2}-2u_x\braket{\widetilde{\overline{f}}_{\alpha}|e_{\alpha x}}+u_x^2\braket{\widetilde{\overline{f}}_{\alpha}|\rho}.
\end{equation}
From Eq.~(\ref{eq:cascadedcollision1}) and substituting for various expressions involving $\braket{\overline{f}_{\alpha}|e_{\alpha x}^m}$,~$\sum_{\beta} \braket{K_{\beta}|e_{\alpha x}^m}\widehat{g}_{\beta}$~and~$\braket{S_{\alpha}|e_{\alpha x}^m}$, where $m=0,1,2$ from the above, yields
\begin{equation}
6\widehat{g}_3+2\widehat{g}_4=\frac{1}{3}\rho-\left(\sum_{\alpha}^{A_3}\right)\otimes \overline{f}_{\alpha}+\rho u_x^2-F_xu_x.
\label{eq:solve3}
\end{equation}

Similarly, from Eq.~(\ref{eq:centralmomenttransformed4})
\begin{equation}
\widehat{\overline{\kappa}}_{yy}^{eq}=1/3\rho=\braket{\widetilde{\overline{f}}_{\alpha}|e_{\alpha y}^2}-2u_y\braket{\widetilde{\overline{f}}_{\alpha}|e_{\alpha y}}+u_y^2\braket{\widetilde{\overline{f}}_{\alpha}|\rho},
\end{equation}
and using $\braket{\overline{f}_{\alpha}|e_{\alpha y}^m}$,~$\sum_{\beta} \braket{K_{\beta}|e_{\alpha y}^m}\widehat{g}_{\beta}$~and~$\braket{S_{\alpha}|e_{\alpha y}^m}$, where $m=0,1,2$ from the above, via the binomial theorem gives
\begin{equation}
6\widehat{g}_3-2\widehat{g}_4=\frac{1}{3}\rho-\left(\sum_{\alpha}^{A_4}\right)\otimes \overline{f}_{\alpha}+\rho u_y^2-F_yu_y.
\label{eq:solve4}
\end{equation}
Solving Eq.~(\ref{eq:solve3}) and (\ref{eq:solve4}) for $\widehat{g}_3$ and $\widehat{g}_4$ yields
\begin{equation}
\widehat{g}_3=\frac{1}{12}\left\{\frac{2}{3}\rho-\left(\sum_{\alpha}^{C_3}+2\sum_{\alpha}^{D_3}\right)\otimes \overline{f}_{\alpha}+\rho (u_x^2+u_y^2)-(F_xu_x+F_yu_y)\right\}, \label{eq:g3}
\end{equation}
and
\begin{equation}
\widehat{g}_4=\frac{1}{4}\left\{\left(\sum_{\alpha}^{E_4}-\sum_{\alpha}^{F_4}\right)\otimes \overline{f}_{\alpha}+\rho (u_x^2-u_y^2)-(F_xu_x-F_yu_y)\right\},\label{eq:g4}
\end{equation}
where
\begin{eqnarray}
C_3&=&\left\{1,2,3,4\right\},\\
D_3&=&\left\{5,6,7,8\right\},\\
E_3&=&\left\{2,4\right\},\\
F_3&=&\left\{1,3\right\}.
\end{eqnarray}
Now, we drop the assumption of equilibration considered above applying relaxation parameters, $\omega_3$ and $\omega_4$, to Eq.~(\ref{eq:g3}) and (\ref{eq:g4}), respectively, to get
\begin{equation}
\widehat{g}_3=\omega_3\frac{1}{12}\left\{-\left(\sum_{\alpha}^{C_3}+2\sum_{\alpha}^{D_3}\right)\otimes \overline{f}_{\alpha}+
\frac{2}{3}\rho+\rho (u_x^2+u_y^2)-(F_xu_x+F_yu_y)\right\}, \label{eq:g3r}
\end{equation}
and
\begin{equation}
\widehat{g}_4=\omega_4\frac{1}{4}\left\{\left(\sum_{\alpha}^{E_4}-\sum_{\alpha}^{F_4}\right)\otimes \overline{f}_{\alpha}+\rho (u_x^2-u_y^2)-(F_xu_x-F_yu_y)\right\}.\label{eq:g4r}
\end{equation}

Let us now consider the central moment $\widehat{\overline{\kappa}}_{xy}^{eq}$ in Eq.~(\ref{eq:centralmomenttransformed5}), i.e.,
\begin{equation}
\widehat{\overline{\kappa}}_{xy}^{eq}=0=\braket{\widetilde{\overline{f}}_{\alpha}|(e_{\alpha x}-u_x)(e_{\alpha y}-u_y)},
\end{equation}
and substituting the expressions for various raw moments, we get
\begin{equation}
\widehat{g}_5=\frac{1}{4}\left\{-\left(\sum_{\alpha}^{A_5}-\sum_{\alpha}^{B_5}\right)\otimes \overline{f}_{\alpha}+\rho u_xu_y-\frac{1}{2}(F_xu_y+F_yu_x)\right\},\label{eq:g5}
\end{equation}
and applying a corresponding relaxation parameter $\omega_5$ to represent over-relaxation for this moment, we obtain,
\begin{equation}
\widehat{g}_5=\omega_5\frac{1}{4}\left\{-\left(\sum_{\alpha}^{A_5}-\sum_{\alpha}^{B_5}\right)\otimes \overline{f}_{\alpha}+\rho u_xu_y-\frac{1}{2}(F_xu_y+F_yu_x)\right\}.\label{eq:g5r}
\end{equation}
It is worth noting that due to a slightly different choice of the basis vector $K_{5}$ for $\ket{e_{\alpha x}e_{\alpha y}}$ from that in~\cite{geier06}, Eq.~(\ref{eq:g5r}) differs from that in~\cite{geier06} by a factor of $-1$ apart from the presence of forcing terms.

We now consider the central moment of the next higher order, i.e. $\widehat{\overline{\kappa}}_{xxy}^{eq}$ in Eq.~(\ref{eq:centralmomenttransformed6}),
$\widehat{\overline{\kappa}}_{xxy}^{eq}=-\frac{1}{6}F_y=\braket{\widetilde{\overline{f}}_{\alpha}|(e_{\alpha x}-u_x)^2(e_{\alpha y}-u_y)}$ and following the procedure as discussed above, we get
\begin{eqnarray}
\widehat{g}_6&=&\frac{1}{4}\left\{\left[\left(\sum_{\alpha}^{A_6}-\sum_{\alpha}^{B_6}\right)
-2u_x\left(\sum_{\alpha}^{A_5}-\sum_{\alpha}^{B_5}\right)-u_y\sum_{\alpha}^{A_3}\right]\otimes \overline{f}_{\alpha}
\right. \nonumber \\
&& \left.
+2\rho u_x^2u_y+\frac{1}{2}(1-u_x^2)F_y-F_xu_xu_y\right\}-2u_x\widehat{g}_5-\frac{1}{2}u_y(3\widehat{g}_3+\widehat{g}_4).\label{eq:g6}
\end{eqnarray}
Notice that $\widehat{g}_6$ depends on $\widehat{g}_{\beta}$, $\beta<6$, which are already post-collision states. So, we relax with relaxation parameter $\omega_6$ only those terms that do no contain these terms, leading to
\begin{eqnarray}
\widehat{g}_6&=&\omega_6\frac{1}{4}\left\{\left[\left(\sum_{\alpha}^{A_6}-\sum_{\alpha}^{B_6}\right)
-2u_x\left(\sum_{\alpha}^{A_5}-\sum_{\alpha}^{B_5}\right)-u_y\sum_{\alpha}^{A_3}\right]\otimes \overline{f}_{\alpha}
\right. \nonumber \\
&& \left.
+2\rho u_x^2u_y+\frac{1}{2}(1-u_x^2)F_y-F_xu_xu_y\right\}-2u_x\widehat{g}_5-\frac{1}{2}u_y(3\widehat{g}_3+\widehat{g}_4), \label{eq:g6r}
\end{eqnarray}
That is, $\widehat{g}_6=\widehat{g}_6(\left\{\overline{f}_{\alpha}\right\},\rho,\overrightarrow{u},\overrightarrow{F},
\widehat{g}_3,\widehat{g}_4,\widehat{g}_5,\omega_6)$.

Considering next, $\widehat{\overline{\kappa}}_{xyy}^{eq}=-\frac{1}{6}F_x=\braket{\widetilde{\overline{f}}_{\alpha}|(e_{\alpha x}-u_x)(e_{\alpha y}-u_y)^2}$ from Eq.~(\ref{eq:centralmomenttransformed7}) and following calculations to transform all the quantities to raw moments, we get
\begin{eqnarray}
\widehat{g}_7&=&\frac{1}{4}\left\{\left[\left(\sum_{\alpha}^{A_7}-\sum_{\alpha}^{B_7}\right)
-2u_y\left(\sum_{\alpha}^{A_5}-\sum_{\alpha}^{B_5}\right)-u_x\sum_{\alpha}^{A_4}\right]\otimes \overline{f}_{\alpha}
\right. \nonumber \\
&& \left.
+2\rho u_xu_y^2+\frac{1}{2}(1-u_y^2)F_x-F_yu_yu_x\right\}-2u_y\widehat{g}_5-\frac{1}{2}u_x(3\widehat{g}_3-\widehat{g}_4),\label{eq:g7}
\end{eqnarray}
Again, notice that $\widehat{g}_7$ depends on $\widehat{g}_{\beta}$, $\beta<6$, which are already post-collision states. So, applying the respective relaxation parameter $\omega_7$ to terms that do no contain them, yields
\begin{eqnarray}
\widehat{g}_7&=&\omega_7\frac{1}{4}\left\{\left[\left(\sum_{\alpha}^{A_7}-\sum_{\alpha}^{B_7}\right)
-2u_y\left(\sum_{\alpha}^{A_5}-\sum_{\alpha}^{B_5}\right)-u_x\sum_{\alpha}^{A_4}\right]\otimes \overline{f}_{\alpha}
\right. \nonumber \\
&& \left.
+2\rho u_xu_y^2+\frac{1}{2}(1-u_y^2)F_x-F_yu_yu_x\right\}-2u_y\widehat{g}_5-\frac{1}{2}u_x(3\widehat{g}_3-\widehat{g}_4),\label{eq:g7r}
\end{eqnarray}
Thus, $\widehat{g}_7=\widehat{g}_7(\left\{\overline{f}_{\alpha}\right\},\rho,\overrightarrow{u},\overrightarrow{F},
\widehat{g}_3,\widehat{g}_4,\widehat{g}_5,\omega_7)$. In other words, $\widehat{g}_{\beta}$ depends on only the lower order moments and not on other components of the same order.

Finally, we consider the central moment of the highest order defined by the discrete particle velocity set (Eq.~(\ref{eq:centralmomenttransformed8})),
$\widehat{\overline{\kappa}}_{xxyy}^{eq}=\frac{1}{9}\rho=\braket{\widetilde{\overline{f}}_{\alpha}|(e_{\alpha x}-u_x)^2(e_{\alpha y}-u_y)^2}$, and apply the procedure as discussed above to transform everything in terms of raw moments to obtain
\begin{eqnarray}
\widehat{g}_8&=&\frac{1}{4}\left\{-\left[\sum_{\alpha}^{A_8}-2u_x\left(\sum_{\alpha}^{A_7}-\sum_{\alpha}^{B_7}\right)
-2u_y\left(\sum_{\alpha}^{A_6}-\sum_{\alpha}^{B_6}\right)
+u_x^2\sum_{\alpha}^{A_4}+u_y^2\sum_{\alpha}^{A_3}+\right.\right.\nonumber\\
&&\left.\left.
4u_xu_y\left(\sum_{\alpha}^{A_5}-\sum_{\alpha}^{B_5}\right)\right]\otimes \overline{f}_{\alpha}
+\frac{1}{9}\rho+3\rho u_x^2u_y^2-(F_xu_xu_y^2+F_yu_yu_x^2)\right\}-2\widehat{g}_3
\nonumber\\
&&-\frac{1}{2}u_x^2(3\widehat{g}_3-\widehat{g}_4)-\frac{1}{2}u_y^2(3\widehat{g}_3+\widehat{g}_4)-4u_xu_y\widehat{g}_5-2u_y\widehat{g}_6-2u_x\widehat{g}_7, \label{eq:g8}
\end{eqnarray}
Clearly, $\widehat{g}_8$ depends on $\widehat{g}_{\beta}$, $\beta<7$, which are already post-collision states and thus, we relax with the parameter $\omega_8$ those terms that do not contain them to finally yield
\begin{eqnarray}
\widehat{g}_8&=&\omega_8\frac{1}{4}\left\{-\left[\sum_{\alpha}^{A_8}-2u_x\left(\sum_{\alpha}^{A_7}-\sum_{\alpha}^{B_7}\right)
-2u_y\left(\sum_{\alpha}^{A_6}-\sum_{\alpha}^{B_6}\right)
+u_x^2\sum_{\alpha}^{A_4}+u_y^2\sum_{\alpha}^{A_3}+\right.\right.\nonumber\\
&&\left.\left.
4u_xu_y\left(\sum_{\alpha}^{A_5}-\sum_{\alpha}^{B_5}\right)\right]\otimes \overline{f}_{\alpha}
+\frac{1}{9}\rho+3\rho u_x^2u_y^2-(F_xu_xu_y^2+F_yu_yu_x^2)\right\}-2\widehat{g}_3
\nonumber\\
&&-\frac{1}{2}u_x^2(3\widehat{g}_3-\widehat{g}_4)-\frac{1}{2}u_y^2(3\widehat{g}_3+\widehat{g}_4)-4u_xu_y\widehat{g}_5-2u_y\widehat{g}_6-2u_x\widehat{g}_7, \label{eq:g8r}
\end{eqnarray}
In order words, $\widehat{g}_8=\widehat{g}_8(\left\{\overline{f}_{\alpha}\right\},\rho,\overrightarrow{u},
\widehat{g}_3,\widehat{g}_4,\widehat{g}_5,\widehat{g}_6,\widehat{g}_7,\omega_8)$. It may be noted that because of a slightly different choice of the basis vector $K_5$, the prefactors for $\widehat{g}_5$ in Eqs.~(\ref{eq:g6r})-(\ref{eq:g8r}) differ from that in~\cite{geier06} by $-1$. Unfortunately, in the seminal work~\cite{geier06}, there are some typographical errors in Eqs.~(20)-(24) of that paper~\cite{geier06} -- in particular, some of the signs in the last lines of its Eq.~(20)-(23), and the expression in the last line of its Eq.~(24) are incorrect.

Thus, the general structure of cascaded collision operator for non-conserved moments may be written as
\begin{equation}
\widehat{g}_{\alpha}=\omega_{\alpha}\left[H_1(\rho,\overrightarrow{u})\star M(\left\{\overline{f}_{\beta}\right\})+H_2(\rho,\overrightarrow{u})\circ N(\overrightarrow{F})\right]+C(\widehat{g}_{\gamma}),
\end{equation}
where $\alpha=3,\ldots,8$,~$\beta=0,1,2,\ldots,8$~and~$\gamma=0,1,2,\ldots,\alpha-1$, and $M$, $N$, $H_1$,~and~$H_2$ represent certain functions, and $\star$ and $\circ$ represent certain operators. On the other hand, in particular, the term $C(\widehat{g}_{\gamma})$ contains the dependence of $\widehat{g}_{\alpha}$ on its corresponding lower order moments leading to a cascaded structure. In other words, cascaded collision operator markedly distinguishes from the SRT and MRT collision operators in that the former is non-commutative. The above derivation involved the choice of a particular form of the central moments of the sources.
In the next section (Sec.~\ref{sec:dealiasinghighercentralsourcemoments}), it will be shown how a different choice could
provide a better representation of its effect on higher order moments.

\section{\label{sec:dealiasinghighercentralsourcemoments}De-aliasing Higher Order Central Source Moments}
Due to the specific formulation of the forcing term employed in Eq.~(\ref{eq:forceansatz}), its corresponding higher order central
moments also have non-zero contributions, even when the fluid is at rest and a homogeneous force is considered. Since they only occur at third
and higher order moments, they do not affect consistency to the Navier-Stokes equations, which emerge at the second-order level
(see Appendix~\ref{app:ChapmanEnskoganalysis}). However, to be conceptually consistent, it is desirable to avoid this effect. Thus,
as a limiting case, we now maintain the effect of the force field only on the components of the first-order central source moments,
and de-alias all the corresponding higher (odd) order central moments, by setting them to zero. That is,
\begin{equation}
\widehat{\Gamma}^{F}_{x^my^n} = \left\{\begin{array}{ll}
   {F_x,}&{m=1,n=0}\\
   {F_y,}&{m=0,n=1}\\
   {0,} &{m+n>1.}
\end{array} \right.
\label{eq:deliasedhighercentralsourcemomentsansatz}
\end{equation}
In effect, the transformed equilibrium central moments $\widehat{\overline{\kappa}}^{eq}_{x^m y^n}$ used in the construction of the
collision operator are modified. Specifically, the third-order transformed equilibrium central moments,
Eqs.~(\ref{eq:centralmomenttransformed6}) and (\ref{eq:centralmomenttransformed7}) now reduce to
\begin{equation}
\widehat{\overline{\kappa}}^{eq}_{xxy}=\widehat{\overline{\kappa}}^{eq}_{xyy}=0,
\end{equation}
while all the other components are the same as before. Moreover, such de-aliasing also modifies the raw moments of the forcing terms
at higher orders. In particular, Eqs.~(\ref{eq:rawmomentsourceterm6})-(\ref{eq:rawmomentsourceterm8}) now become
\begin{eqnarray}
\braket{S_{\alpha}|e_{\alpha x}^2e_{\alpha y}}&=&\sum_{\alpha}S_{\alpha} e_{\alpha x}^2e_{\alpha
y}=F_yu_x^2+2F_xu_xu_y,\label{eq:dealiasedrawmomentsourceterm6}\\
\braket{S_{\alpha}|e_{\alpha x}e_{\alpha y}^2}&=&\sum_{\alpha}S_{\alpha} e_{\alpha x}e_{\alpha
y}^2=F_xu_y^2+2F_yu_yu_x,\label{eq:dealiasedrawmomentsourceterm7}\\
\braket{S_{\alpha}|e_{\alpha x}^2e_{\alpha y}^2}&=&\sum_{\alpha}S_{\alpha} e_{\alpha x}^2e_{\alpha
y}^2=2F_xu_xu_y^2+2F_yu_yu_x^2.\label{eq:dealiasedrawmomentsourceterm8}
\end{eqnarray}
while the lower order moments remain unaltered. Notice that terms such as $1/3F_x$ and $1/3F_y$ do not anymore appear in the
third-order source moments, while $2/3F_xu_x$ and $2/3F_yu_y$ are eliminated from the fourth-order source moments as a result of the
use of de-aliased central source moments (Eq.~(\ref{eq:deliasedhighercentralsourcemomentsansatz})). Hence, when the fluid is rest, the
force fields do not influence the third and higher order raw source moments, which is physically consistent.

The computation of the source terms in velocity space $S_{\alpha}$ using Eqs.~(\ref{eq:vsourceterms0})-(\ref{eq:vsourceterms8}), which involve $\widehat{m}^{s}_{\beta}$, are also naturally influenced by the above changes. In this regard, while
$\widehat{m}^{s}_{\beta}$, for $\beta=0,1,2,\ldots,5$ remain unmodified, the higher order moments for $\beta=6,7,8$ are altered. The
expressions for these latter quantities now become
\begin{eqnarray}
\widehat{m}^{s}_{6}=\braket{K_6|S_{\alpha}}&=&(2-3u_x^2)F_y-6F_xu_xu_y, \label{eq:dealiasedsourceterm6}\\
\widehat{m}^{s}_{7}=\braket{K_7|S_{\alpha}}&=&(2-3u_y^2)F_x-6F_yu_yu_x, \label{eq:dealiasedsourceterm7}\\
\widehat{m}^{s}_{8}=\braket{K_8|S_{\alpha}}&=&6\left[(3u_y^2-2)F_xu_x+(3u_x^2-2)F_yu_y\right]. \label{eq:dealiasedsourceterm8}
\end{eqnarray}

The cascaded collision operator can now be constructed using the procedure presented in Sec.~\ref{sec:cascadedcollisionforcing}. The
use of modified source moments do not alter the collision kernel corresponding to $\widehat{g}_{\beta}$, where $\beta=0,1,2,\ldots,5$
and $\beta=8$. They are the same as those presented in Sec.~\ref{sec:cascadedcollisionforcing}. On the other hand, the third-order
collision kernel contributions are modified, which are now summarized as follows:
\begin{eqnarray}
\widehat{g}_6&=&\omega_6\frac{1}{4}\left\{\left[\left(\sum_{\alpha}^{A_6}-\sum_{\alpha}^{B_6}\right)
-2u_x\left(\sum_{\alpha}^{A_5}-\sum_{\alpha}^{B_5}\right)-u_y\sum_{\alpha}^{A_3}\right]\otimes \overline{f}_{\alpha}
\right. \nonumber \\
&& \left.
+2\rho u_x^2u_y-\frac{1}{2}u_x^2F_y-F_xu_xu_y\right\}-2u_x\widehat{g}_5-\frac{1}{2}u_y(3\widehat{g}_3+\widehat{g}_4),
\label{eq:dealiasedg6r}
\end{eqnarray}
and
\begin{eqnarray}
\widehat{g}_7&=&\omega_7\frac{1}{4}\left\{\left[\left(\sum_{\alpha}^{A_7}-\sum_{\alpha}^{B_7}\right)
-2u_y\left(\sum_{\alpha}^{A_5}-\sum_{\alpha}^{B_5}\right)-u_x\sum_{\alpha}^{A_4}\right]\otimes \overline{f}_{\alpha}
\right. \nonumber \\
&& \left.
+2\rho u_xu_y^2-\frac{1}{2}u_y^2F_x-F_yu_yu_x\right\}-2u_y\widehat{g}_5-\frac{1}{2}u_x(3\widehat{g}_3-\widehat{g}_4).
\label{eq:dealiasedg7r}
\end{eqnarray}
Again, evidently, when the fluid is at rest, the force fields do not have direct influence on $\widehat{g}_6$ and $\widehat{g}_7$.
Thus, the above formulation eliminates spurious effects resulting from forcing due to the finiteness of the lattice set for higher order moments, similar to
that by Guo \emph{et al}.~\cite{guo02} for other LBM approaches. Indeed, a Chapman-Enskog multiscale moment expansion analysis
carried out in Appendix~\ref{app:ChapmanEnskoganalysis} will establish the consistency of this special formulation of the
central moments based LBM to the desired macroscopic fluid flow equations.                                                                                                                               The shear and bulk kinematic viscosities is found to be dependent on the relaxation parameters
$\omega_3=\omega^\chi$~and~$\omega_4=\omega_5=\omega^\nu$, respectively. In particular, the shear
viscosity satisfies $\nu=c_s^2\left(\frac{1}{\omega^\nu}-\frac{1}{2}\right)$. The rest of the
relaxation parameters in this MRT cascaded formulation can be tuned to maintain numerical stability. One
particular choice suggested by Geier is to equilibrate higher order, in particular, the third-order moments,
$\omega_6=\omega_7=\omega_8=1$~\cite{geier08b}. Other possible choices could be also considered that involve
over-relaxation of these moments at certain carefully selected relaxation rates so as to control numerical
dissipation while maintaining computational stability. On the other hand, as shown in
Appendix~\ref{app:GeneralizedEqilibriumSources}, when the central moments based LBM as derived in this work
is executed as a MRT cascaded process it implies generalization of both equilibrium and sources in the
lattice frame reference which also depend on the ratio of various relaxation times. However, it does not affect
the overall consistency of the approach to the macroscopic equations as it influences only higher order contributions.
The discussions so far considered the cascaded collision operator to be explicit in time.
Appendix~\ref{app:TimeImplicitnessCascadedCollision} presents with the possibility of introducing time-implicitness
in the cascaded collision operator.

\section{\label{sec:computationalprocedure}Computational Procedure}
The main element of the computational procedure consists of performing the cascaded collision, including the forcing terms, i.e. Eq.~(\ref{eq:cascadedcollision1}) along with Eq.~(\ref{eq:cascadedcollisionoperator}), which can be expanded as follows:
\begin{eqnarray}
\widetilde{\overline{f}}_{0}&=&\overline{f}_{0}+\left[\widehat{g}_0-4(\widehat{g}_3-\widehat{g}_8)\right]+S_0, \label{eq:postcollisionf0}\\
\widetilde{\overline{f}}_{1}&=&\overline{f}_{1}+\left[\widehat{g}_0+\widehat{g}_1-\widehat{g}_3+\widehat{g}_4    +2(\widehat{g}_7-\widehat{g}_8)\right]+S_1, \label{eq:postcollisionf1}\\
\widetilde{\overline{f}}_{2}&=&\overline{f}_{2}+\left[\widehat{g}_0+\widehat{g}_2-\widehat{g}_3-\widehat{g}_4
+2(\widehat{g}_6-\widehat{g}_8)\right]+S_2, \label{eq:postcollisionf2}\\
\widetilde{\overline{f}}_{3}&=&\overline{f}_{3}+\left[\widehat{g}_0-\widehat{g}_1-\widehat{g}_3+\widehat{g}_4
-2(\widehat{g}_7+\widehat{g}_8)\right]+S_3, \label{eq:postcollisionf3}\\
\widetilde{\overline{f}}_{4}&=&\overline{f}_{4}+\left[\widehat{g}_0-\widehat{g}_2-\widehat{g}_3-\widehat{g}_4
-2(\widehat{g}_6+\widehat{g}_8)\right]+S_4, \label{eq:postcollisionf4}\\
\widetilde{\overline{f}}_{5}&=&\overline{f}_{5}+\left[\widehat{g}_0+\widehat{g}_1+\widehat{g}_2+2\widehat{g}_3
+\widehat{g}_5-\widehat{g}_6-\widehat{g}_7+\widehat{g}_8\right]+S_5, \label{eq:postcollisionf5}\\
\widetilde{\overline{f}}_{6}&=&\overline{f}_{6}+\left[\widehat{g}_0-\widehat{g}_1+\widehat{g}_2+2\widehat{g}_3
-\widehat{g}_5-\widehat{g}_6+\widehat{g}_7+\widehat{g}_8\right]+S_6, \label{eq:postcollisionf6}\\
\widetilde{\overline{f}}_{7}&=&\overline{f}_{7}+\left[\widehat{g}_0-\widehat{g}_1-\widehat{g}_2+2\widehat{g}_3
+\widehat{g}_5+\widehat{g}_6+\widehat{g}_7+\widehat{g}_8\right]+S_7, \label{eq:postcollisionf7}\\
\widetilde{\overline{f}}_{8}&=&\overline{f}_{8}+\left[\widehat{g}_0+\widehat{g}_1-\widehat{g}_2+2\widehat{g}_3
-\widehat{g}_5+\widehat{g}_6-\widehat{g}_7+\widehat{g}_8\right]+S_8. \label{eq:postcollisionf8}
\end{eqnarray}
Here, the terms $\widehat{g}_{\beta}$ can be obtained in a sequential manner, i.e. evolving towards higher moment orders from Eqs.~(\ref{eq:g3r}),~(\ref{eq:g4r}),~(\ref{eq:g5r}),~(\ref{eq:g6r}),~(\ref{eq:g7r}),~and~(\ref{eq:g8r}). It consists of terms that involve summation of $\overline{f}_{\alpha}$ over various subsets of the particle velocity set. The source terms
$S_{\beta}$ are computed from Eqs.~(\ref{eq:vsourceterms0})-(\ref{eq:vsourceterms8}). Once the post-collision values, i.e. $\widetilde{\overline{f}}_{\alpha}$ are known, the streaming step can be performed in the usual manner to obtain the updated value of $\overline{f}_{\alpha}$ (Eq.~(\ref{eq:cascadedstreaming1})). Subsequently, the hydrodynamic fields, viz., the local fluid density and velocity can be computed from Eqs.~(\ref{eq:densitycalculation}) and (\ref{eq:velocitycalculation}), respectively. Depending on the
specific choice of the ansatz for the central source moments, appropriate expressions for $\widehat{g}_{\beta}$ and $\widehat{m}_{\beta}^{s}$ need to be used (see Secs.~\ref{sec:cascadedcollisionforcing} and
\ref{sec:dealiasinghighercentralsourcemoments}). In the above procedure, careful optimization needs to be carried out
to reduce the number of floating-point operations.

\section{\label{sec:compexperiments}Computational Experiments}
\begin{figure}
\includegraphics[width = 90mm, angle = 270]{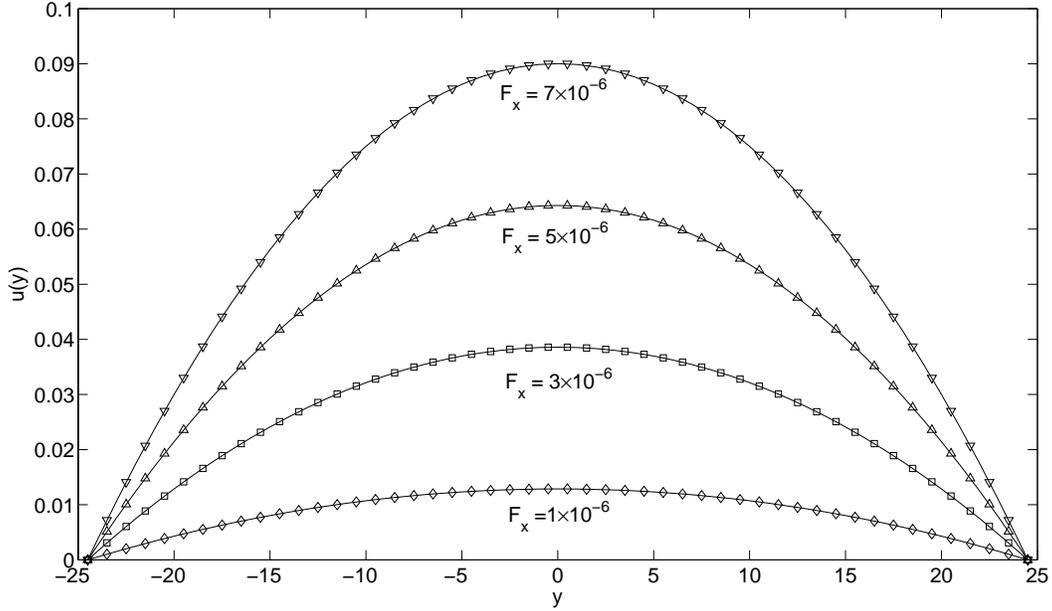}
\caption{\label{fig:Poiseuille} Flow between parallel plates with constant body force:
Comparison of velocity profiles computed by Cascaded-LBM with forcing term (symbols) with analytical solution (lines) for different values of the body force $F_x$.}
\end{figure}

In order to validate the numerical accuracy of the new computational approach presented in this work, we performed simulations
for canonical fluid flow problems subjected to different types of forces, where analytical solutions are available. We will now present results obtained by employing the Cascaded-LBM with de-aliased higher order source central moments (as discussed in
Sec.~\ref{sec:dealiasinghighercentralsourcemoments}), which will be compared with corresponding analytical solutions. The first problem considered is the flow between parallel plates subjected to a constant body force. We considered $3\times 51$ lattice nodes to resolve the computational domain, where periodic boundary conditions are imposed in the flow direction and the no slip boundary condition at the walls is represented by means of the standard link bounce back technique. The relaxation parameters are given such that $\omega_4=\omega_5=1.754$, while the remaining ones are set to unity and the computations are performed for different values of the component of the body force in the flow direction, i.e. $F_x$ with $F_y=0$. Figure~\ref{fig:Poiseuille} shows a comparison of the computed velocity profiles with the standard analytical solution (Poiseuille's parabolic profile, with the maximum velocity $u_0=F_xL^2/(2\nu)$, where $L$ is the half-width between the plates and $\nu$ is the fluid's kinematic viscosity) for different values of $F_x$. Excellent agreement is seen.
In order to quantify the difference between the computed and analytical solution, the relative global error given in terms of
the Euclidean (second) norm is presented in Table I. Thus, for the above given set of parameters and resolution, it is $O(10^{-4})$.
\begin{table}[h]
\label{tab:errorresidual}
  \begin{tabular}{| c | c | }
    \hline
    Magnitude of body force ($F_x$) & Relative global error ($||\delta u||_2$) \\ \hline \hline
    $1 \times 10^{-6}$ & $3.999\times 10^{-4}$ \\
    $3 \times 10^{-6}$ & $3.895\times 10^{-4}$ \\
    $5 \times 10^{-6}$ & $3.837\times 10^{-4}$ \\
    $7 \times 10^{-6}$ & $3.839\times 10^{-4}$ \\
    \hline
    \end{tabular}\caption{Relative global error for the Poiseuille flow problem. $||\delta u||_2=\sum_i||(u_{c,i}-u_{a,i})||_2 / \sum_i ||u_{a,i}||_2$, where $u_{c,i}$ and $u_{a,i}$ are
    computed and analytical solutions, respectively, and the summation is over the entire domain.}
\end{table}

\begin{figure}
\includegraphics[width = 90mm, angle = 270]{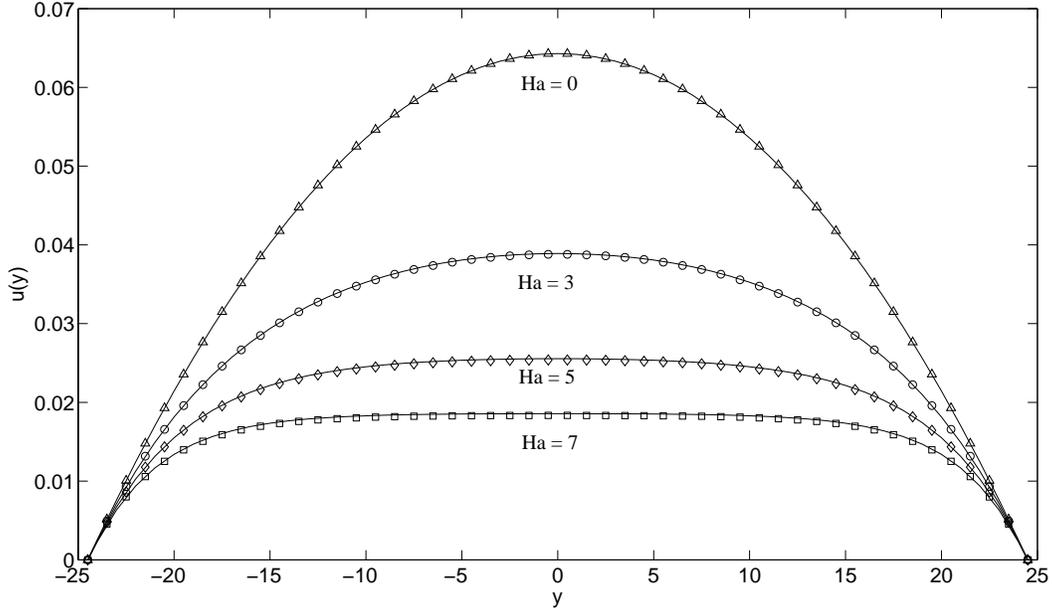}
\caption{\label{fig:Hartmann} Flow between parallel plates with a spatially varying body force:
Comparison of velocity profiles computed by Cascaded-LBM with forcing term (symbols) with analytical solution (lines) for prescribed Lorentz force at different Hartmann numbers.}
\end{figure}

The second problem considered involves a spatially varying body force. One classical problem in this regard is the Hartmann flow, i.e. flow between parallel plates subjected to a magnetic field $B_y=B_0$ imposed in the perpendicular direction to the fluid motion. If
$F_b$ is the driving force of the fluid due to imposed pressure gradient and $\mathrm{Ha}$ is the Hartmann number that characterizes the ratio of force due to magnetic field and the viscous force, then the induced magnetic field in the flow direction $B_x$ is given by $B_x=\frac{F_bL}{B_0}\left[\frac{sinh\left(\mathrm{Ha}\frac{y}{L}\right)}{sinh(\mathrm{Ha})}-\frac{y}{L}\right]$, where the coordinate distance $y$ is measured from a position equidistant between the plates. The interaction of the flow field with the magnetic field results in a variable retarding force $F_{mx}=B_y\frac{dB_x}{dy}$ and $F_{my}=-B_x\frac{dB_x}{dy}$, and, in turn, the net force acting on the fluid is $F_x=F_b+F_{mx}$ and $F_y=F_{my}$. We considered the same number of lattice nodes and the same values of the relaxation parameters as before, with $F_b=5\times 10^{-6}$ and $B_0=8\times 10^{-3}$ and varied the values of $\mathrm{Ha}$. The analytical solution for this problem is $u_x=\frac{F_bL}{B_0}\sqrt{\frac{\eta}{\nu}}coth(\mathrm{Ha})\left[1-\frac{cosh\left(\mathrm{Ha}\frac{y}{L}\right)}{cosh(\mathrm{Ha})}\right]$, where the magnetic resistivity $\eta$ is related to $\mathrm{Ha}$ through $\eta=\frac{B_0^2L^2}{\mathrm{Ha}^2\nu}$. The computed velocity profiles are compared with the analytical solution for different values of $\mathrm{Ha}$ in Fig.~\ref{fig:Hartmann}.
As expected, the velocity profiles become more flattened with increasing values of $\mathrm{Ha}$, while the case with $\mathrm{Ha}=0$ reduces to the earlier problem. The computed velocity profiles are found to agree very well with the analytical results. The relative
global errors for this problem are presented in Table II. It can be seen that they are dependent on the value of $\mathrm{Ha}$ when the same grid resolution is used for different cases.
In particular, the relative error increases as the value of $\mathrm{Ha}$ is increased for the same resolution. This can be explained
as follows. This flow problem is characterized by the presence of boundary layers -- the Hartmann layers -- whose
thickness is inversely proportional to $\sqrt{\mathrm{Ha}}$. That is, the Hartmann layer becomes thinner as the value of $\mathrm{Ha}$ is increased. Thus, resolution of this boundary layer would require increasingly more number nodes that are clustered near
walls as $\mathrm{Ha}$ is increased to maintain the same accuracy. Otherwise, when the same number of grid nodes that are uniformly
distributed is employed, the relatively error norm is expected to increase with $\mathrm{Ha}$. Indeed, local grid refinement
employing a suitable boundary layer transformation can maintain similar accuracy for different $\mathrm{Ha}$ as was done with
other LBM formulations recently~\cite{pattison08}. Extension of the local grid refinement approaches for the central moment based LBM to resolve boundary layers and sharp gradients in solutions are subjects of future studies.
\begin{table}[h]
\label{tab:errorresidualHa}
  \begin{tabular}{| c | c | }
    \hline
    Hartmann number ($\mathrm{Ha}$) & Relative global error ($||\delta u||_2$) \\ \hline \hline
    $0.0$ & $3.837\times 10^{-4}$ \\
    $3.0$ & $2.140\times 10^{-3}$ \\
    $5.0$ & $5.967\times 10^{-3}$ \\
    $7.0$ & $1.091\times 10^{-2}$ \\
    \hline
    \end{tabular}\caption{Relative global error for the Hartmann flow problem. $||\delta u||_2=\sum_i||(u_{c,i}-u_{a,i})||_2 /
    \sum_i ||u_{a,i}||_2$, where $u_{c,i}$ and $u_{a,i}$ are
    computed and analytical solutions, respectively, and the summation is over the entire domain.}
\end{table}

\begin{figure}
\includegraphics[width = 90mm, angle = 270]{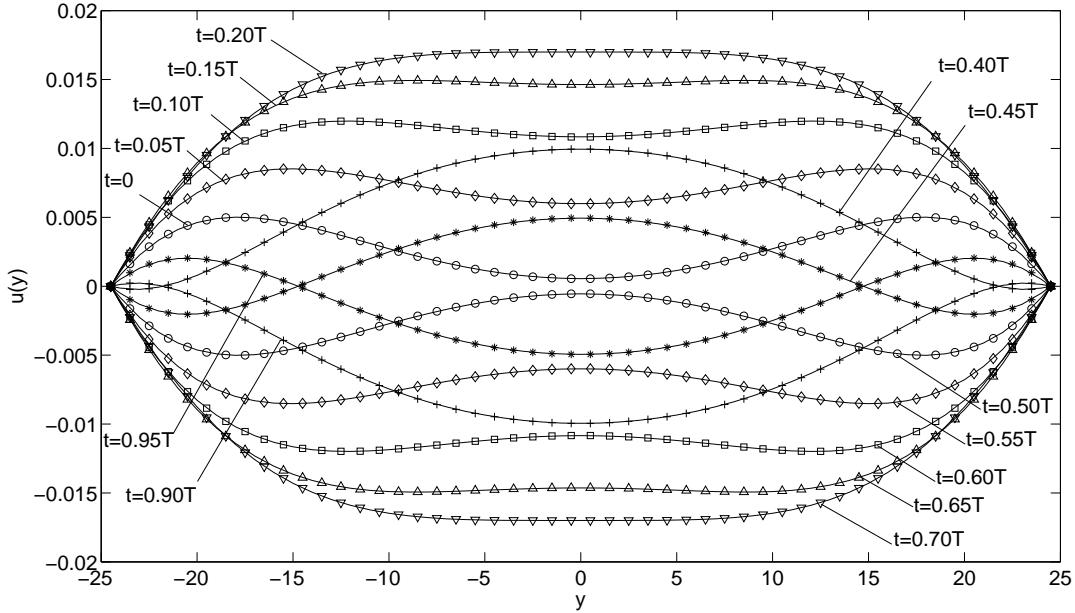}
\caption{\label{fig:Womersley} Flow between parallel plates with a temporally varying body force:
Comparison of velocity profiles computed by Cascaded-LBM with forcing term (symbols) with analytical solution (lines) at different instants within a time period $T$.}
\end{figure}

The last problem that we considered involves a temporally varying body force. An important canonical problem in this regard is the
flow between two parallel plates driven by a force sinusoidally varying in time. That is, we considered $F_x=F_bcos(\omega t)$, where $F_b$ is the peak value of the applied force, while $\omega_p=2\pi/T$ is the angular frequency where $T$ is the time period. This problem is characterized by $\mathrm{Wo}=\sqrt{\frac{\omega_p}{\nu}}L$, a dimensionless number arising from its original analysis by Womersley. The analytical velocity profile for this flow is $u_x=\mathcal{R}\left[\frac{iF_b}{\omega_p}\left\{1-\frac{cos\left(\gamma\frac{y}{L}\right)}{cos(\gamma)}\right\}e^{i\omega_pt}\right]$,
where $\gamma=\sqrt{-i\mathrm{Wo}^2}$. We considered $F_b=1\times 10^{-5}$ and $\mathrm{Wo}=12.71$, while maintaining the number of lattice nodes and the values of the relaxation parameters to be same as in the first problem. Figure~\ref{fig:Womersley} shows a comparison of the computed velocity profiles with analytical solution for different instants within the duration of the time period $T$ of the cycle.
Evidently, the new computational approach is able to reproduce the complex flow features for this problem involving the presence of
Stokes layer very well. Table III presents the relative global errors at different instants within the time period $T$, corresponding to those in Fig.~\ref{fig:Womersley}. The relatively differences between computed and analytical solutions vary between different time instants. On the other hand, they are identical for instants shifted by the half time period implying that the computations are able to reproduce temporal variations without any time lag as compared with analytical solutions.
\begin{table}[h]
\label{tab:errorresidualWo}
  \begin{tabular}{| c | c | }
    \hline
    Time instant ($t$) & Relative global error ($||\delta u||_2$) \\ \hline \hline
    $0$     & $4.195\times 10^{-3}$ \\                                                                                                $0.05T$ & $1.701\times 10^{-3}$ \\
    $0.10T$ & $1.060\times 10^{-3}$ \\
    $0.15T$ & $7.548\times 10^{-4}$ \\
    $0.20T$ & $5.906\times 10^{-4}$ \\
    $0.40T$ & $1.842\times 10^{-3}$ \\
    $0.45T$ & $4.611\times 10^{-4}$ \\
    $0.50T$ & $4.195\times 10^{-3}$ \\
    $0.55T$ & $1.701\times 10^{-3}$ \\
    $0.60T$ & $1.060\times 10^{-3}$ \\
    $0.65T$ & $7.548\times 10^{-4}$ \\
    $0.70T$ & $5.906\times 10^{-3}$ \\
    $0.90T$ & $1.842\times 10^{-3}$ \\
    $0.95T$ & $4.611\times 10^{-3}$ \\
    \hline
    \end{tabular}\caption{Relative global error for the Womersley flow problem.
    $||\delta u||_2(t)=\sum_i||(u_{c,i}(t)-u_{a,i}(t))||_2 /
    \sum_i ||u_{a,i}(t)||_2$, where $u_{c,i}(t)$ and $u_{a,i}(t)$ are
    computed and analytical solutions, respectively, at instant $t$ within a time period $T$ and the summation is over the entire domain.}
\end{table}

It may be noted that for all the three benchmark problems presented above, essentially same numerical results are obtained when the
de-aliasing in the forcing is turned off, i.e. expressions presented in Sec.~\ref{sec:cascadedcollisionforcing} is used.
This is because both forms differ only in third and higher orders, while they are both consistent at the second order
level with the Navier-Stokes equations, from which the analytical solutions are derived. It would be interesting to
carry out detailed numerical error analysis as well as stability analysis of the central moment based LBM for different grid resolutions and characteristic parameters, and for various canonical flow problems in future investigations.

\section{\label{sec:summaryconclusions}Summary and Conclusions}
In this paper, we discussed a systematic procedure for the derivation of forcing terms based on the central moments in the Cascaded-LBM. The main elements involved in this regard are the binomial theorem that relates the central moments and raw moments of various orders and the associated orthogonal properties. The discrete source terms are obtained by matching with the corresponding continuous central moment of a given
order. For the latter, we consider an ansatz based on the local Maxwell distribution. Its variant involving a de-aliased higher order central source moments, which recovers physically consistent higher order effects when the fluid is at rest, is also derived.
Effectively explicit and temporally second-order forms of forcing terms are obtained through a transformation of the distribution function, which contributes to the cascaded collision. When the values of the free parameters in the continuous equilibrium (Maxwell) distribution, i.e. speed of sound and those in the orthogonalization process of the moment basis from the discrete velocity set are
chosen, they completely determine the various coefficients of both the cascaded collision operator and the source terms. The equilibrium distribution and the source terms in velocity space are proper polynomials and contain higher order terms.
By construction, the source terms are Galilean invariant. It is found that both the equilibrium and source terms generalize
when the cascaded formulation is represented as a relaxation process in the lattice frame of reference. While the Cascaded-LBM
with forcing terms is based on a frame invariant kinetic theory, its consistency to the Navier-Stokes equations is shown by means of a Chapman-Enskog moment expansion analysis. It is found that the new approach reproduces analytical solutions for canonical problems that involve either constant or spatially or temporally varying forces with excellent quantitative accuracy. The approach presented in this paper can be extended to other types of lattices such as the D3Q27 model in three dimensions~\cite{premnath09b}.

\appendix

\section{\label{app:ChapmanEnskoganalysis}Chapman-Enskog Multiscale Analysis}
In this section, let us perform a Chapman-Enskog analysis of the central moment formulation of the LBM using the consistent forcing
terms derived in Sec.~\ref{sec:dealiasinghighercentralsourcemoments}. For ease of presentation and analysis, we will make a
particular assumption regarding the collision operator in this section. It will then be pointed out in the next section that
relaxing such assumption amounting to the use of fully coherent cascaded collision kernel does not affect the consistency analysis
presented here. First, some preliminaries are provided. In particular, we define a transformation matrix corresponding to the
following ``nominal'' moment basis on which the analysis is performed:
\begin{equation}
\mathcal{T}=\left[\ket{\rho},\ket{e_{\alpha x}},\ket{e_{\alpha y}},\ket{e_{\alpha x}^2+e_{\alpha y}^2},\ket{e_{\alpha x}^2-e_{\alpha y}^2},\ket{e_{\alpha x}e_{\alpha y}},\ket{e_{\alpha x}^2e_{\alpha y}},\ket{e_{\alpha x}e_{\alpha y}^2},\ket{e_{\alpha x}^2e_{\alpha y}^2}\right],
\label{eq:nominalmomentbasis}
\end{equation}
It is convenient to carry out the multiscale expansion in terms of various raw moments.
Thus, we also define the following \emph{raw} moments, where the superscript ``prime'' symbol is used here
and henceforth to designate that the moment is of raw type:
\begin{eqnarray}
\widehat{\kappa}_{x^m y^n}^{'}&=&\sum_{\alpha}f_{\alpha}e_{\alpha x}^me_{\alpha y}^n=\braket{e_{\alpha x}^m e_{\alpha y}^n|f_{\alpha}},\\
\widehat{\sigma}_{x^m y^n}^{'}&=&\sum_{\alpha}S_{\alpha}e_{\alpha x}^me_{\alpha y}^n=\braket{e_{\alpha x}^m e_{\alpha y}^n|S_{\alpha}},\\
\widehat{\kappa}_{x^m y^n}^{eq'}&=&\sum_{\alpha}f_{\alpha}^{eq}e_{\alpha x}^me_{\alpha y}^n=\braket{e_{\alpha x}^m e_{\alpha y}^n|f_{\alpha}^{eq}},\\
\widehat{\overline{\kappa}}_{x^m y^n}^{'}&=&\sum_{\alpha}\overline{f}_{\alpha}e_{\alpha x}^me_{\alpha y}^n=\braket{e_{\alpha x}^m e_{\alpha y}^n|\overline{f}_{\alpha}},\\
\widehat{\overline{\kappa}}_{x^m y^n}^{eq'}&=&\sum_{\alpha}\overline{f}_{\alpha}^{eq}e_{\alpha x}^me_{\alpha y}^n=\braket{e_{\alpha x}^m e_{\alpha y}^n|\overline{f}_{\alpha}^{eq}}.
\end{eqnarray}
It follows that $\widehat{\overline{\kappa}}_{x^m y^n}^{'}=\widehat{\kappa}_{x^m y^n}^{'}-\frac{1}{2}\widehat{\sigma}_{x^m y^n}^{'}$ and $\widehat{\overline{\kappa}}_{x^m y^n}^{eq'}=\widehat{\kappa}_{x^m y^n}^{eq'}-\frac{1}{2}\widehat{\sigma}_{x^m y^n}^{'}$.

We now re-write various different central moments in terms of their corresponding raw moments by applying the binomial theorem. First, the non-conserved part of the central moments can be written as functions of various raw moments as follows:
\begin{eqnarray}
\widehat{\overline{\kappa}}_{xx}&=&\widehat{\overline{\kappa}}_{xx}^{'}-\rho u_x^2+F_xu_x,\\
\widehat{\overline{\kappa}}_{yy}&=&\widehat{\overline{\kappa}}_{yy}^{'}-\rho u_y^2+F_yu_y,\\
\widehat{\overline{\kappa}}_{xy}&=&\widehat{\overline{\kappa}}_{xy}^{'}-\rho u_x u_y+\frac{1}{2}(F_xu_y+F_yu_x),\\
\widehat{\overline{\kappa}}_{xxy}&=&\widehat{\overline{\kappa}}_{xxy}^{'}-2u_x\widehat{\overline{\kappa}}_{xy}^{'}
                                    -u_y\widehat{\overline{\kappa}}_{xx}^{'}+2\rho u_x^2 u_y-\frac{1}{2}F_yu_x^2-F_xu_xu_y,\\
\widehat{\overline{\kappa}}_{xyy}&=&\widehat{\overline{\kappa}}_{xyy}^{'}-2u_y\widehat{\overline{\kappa}}_{xy}^{'}
                                    -u_x\widehat{\overline{\kappa}}_{yy}^{'}+2\rho u_x u_y^2-\frac{1}{2}F_xu_y^2-F_yu_yu_x,\\
\widehat{\overline{\kappa}}_{xxyy}&=&\widehat{\overline{\kappa}}_{xxyy}^{'}-2u_x\widehat{\overline{\kappa}}_{xyy}^{'}
                                    -2u_y\widehat{\overline{\kappa}}_{xxy}^{'}+u_x^2\widehat{\overline{\kappa}}_{yy}^{'}
                                    +u_y^2\widehat{\overline{\kappa}}_{xx}^{'}+4u_xu_y\widehat{\overline{\kappa}}_{xy}^{'}\nonumber\\
                                    &&-3\rho u_x^2 u_y^2+F_xu_xu_y^2+F_yu_yu_x^2.
\end{eqnarray}
The raw moments of the equilibrium distribution and source terms of various order are:
\begin{eqnarray}
\widehat{\kappa}^{eq'}_{0}&=&\rho, \\
\widehat{\kappa}^{eq'}_{x}&=&\rho u_x, \\
\widehat{\kappa}^{eq'}_{y}&=&\rho u_y, \\
\widehat{\kappa}^{eq'}_{xx}&=&\frac{1}{3}\rho+\rho u_x^2,\\
\widehat{\kappa}^{eq'}_{yy}&=&\frac{1}{3}\rho+\rho u_y^2,\\
\widehat{\kappa}^{eq'}_{xy}&=&\rho u_x u_y, \\
\widehat{\kappa}^{eq'}_{xxy}&=&\frac{1}{3}\rho u_y+\rho u_x^2u_y, \\
\widehat{\kappa}^{eq'}_{xyy}&=&\frac{1}{3}\rho u_x+\rho u_xu_y^2, \\
\widehat{\kappa}^{eq'}_{xxyy}&=&\frac{1}{9}\rho+\frac{1}{3}\rho (u_x^2+u_y^2)+\rho u_x^2u_y^2,
\end{eqnarray}
and
\begin{eqnarray}
\widehat{\sigma}^{'}_{0}&=&0, \\
\widehat{\sigma}^{'}_{x}&=&F_x, \\
\widehat{\sigma}^{'}_{y}&=&F_y, \\
\widehat{\sigma}^{'}_{xx}&=&2F_xu_x,\\
\widehat{\sigma}^{'}_{yy}&=&2F_yu_y,\\
\widehat{\sigma}^{'}_{xy}&=&F_xu_y+F_yu_x, \\
\widehat{\sigma}^{'}_{xxy}&=&F_yu_x^2+2F_xu_xu_y, \\
\widehat{\sigma}^{'}_{xyy}&=&F_xu_y^2+2F_yu_yu_x, \\
\widehat{\sigma}^{'}_{xxyy}&=&2F_xu_xu_y^2+2F_yu_yu_x^2,
\end{eqnarray}
respectively.

In the above notation, the cascaded collision kernel may be more compactly written as
\begin{eqnarray}
\widehat{g}_3&=&\frac{\omega_3}{12}\left\{ \frac{2}{3}\rho+\rho(u_x^2+u_y^2)
-(\widehat{\overline{\kappa}}_{xx}^{'}+\widehat{\overline{\kappa}}_{yy}^{'})
-\frac{1}{2}(\widehat{\sigma}_{xx}^{'}+\widehat{\sigma}_{yy}^{'})
\right\}, \label{eq:collisionkernelcompact3}\\
\widehat{g}_4&=&\frac{\omega_4}{4}\left\{\rho(u_x^2-u_y^2)
-(\widehat{\overline{\kappa}}_{xx}^{'}-\widehat{\overline{\kappa}}_{yy}^{'})
-\frac{1}{2}(\widehat{\sigma}_{xx}^{'}-\widehat{\sigma}_{yy}^{'})
\right\}, \label{eq:collisionkernelcompact4}\\
\widehat{g}_5&=&\frac{\omega_5}{4}\left\{\rho u_x u_y
-\widehat{\overline{\kappa}}_{xy}^{'}
-\frac{1}{2}\widehat{\sigma}_{xy}^{'}
\right\}, \label{eq:collisionkernelcompact5}\\
\widehat{g}_6&=&\frac{\omega_6}{4}\left\{2\rho u_x^2 u_y+\widehat{\overline{\kappa}}_{xxy}^{'}
              -2u_x\widehat{\overline{\kappa}}_{xy}^{'}-u_y\widehat{\overline{\kappa}}_{xx}^{'}-\frac{1}{2}\widehat{\sigma}_{xxy}
              \right\}-\frac{1}{2}u_y(3\widehat{g}_3+\widehat{g}_4)-2u_x\widehat{g}_5, \label{eq:collisionkernelcompact6}\\
\widehat{g}_7&=&\frac{\omega_7}{4}\left\{2\rho u_x u_y^2+\widehat{\overline{\kappa}}_{xyy}^{'}
              -2u_y\widehat{\overline{\kappa}}_{xy}^{'}-u_x\widehat{\overline{\kappa}}_{yy}^{'}-\frac{1}{2}\widehat{\sigma}_{xyy}
              \right\}-\frac{1}{2}u_x(3\widehat{g}_3-\widehat{g}_4)-2u_y\widehat{g}_5, \label{eq:collisionkernelcompact7}\\
\widehat{g}_8&=&\frac{\omega_8}{4}\left\{\frac{1}{9}\rho+3\rho u_x^2 u_y^2-\left[\widehat{\overline{\kappa}}_{xxyy}^{'}
                                 -2u_x\widehat{\overline{\kappa}}_{xyy}^{'}-2u_y\widehat{\overline{\kappa}}_{xxy}^{'}
                                 +u_x^2\widehat{\overline{\kappa}}_{yy}^{'}+u_y^2\widehat{\overline{\kappa}}_{xx}^{'}\right.\right.
                                 \nonumber \\
                                 &&\left.\left.+4u_xu_y\widehat{\overline{\kappa}}_{xy}^{'}
                                 \right]-\frac{1}{2}\widehat{\sigma}_{xxyy}^{'}
                                  \right\}-2\widehat{g}_3-\frac{1}{2}u_y^2(3\widehat{g}_3+\widehat{g}_4)
                                  -\frac{1}{2}u_x^2(3\widehat{g}_3-\widehat{g}_4)\nonumber\\
                                  &&-4u_xu_y\widehat{g}_5-2u_y\widehat{g}_6
                                  -2u_x\widehat{g}_7.\label{eq:collisionkernelcompact8}
\end{eqnarray}

Instead of considering the above collision operator, for now, in what follows, let us specialize the collision term. In this regard,
we first re-write the cascaded collision step, Eq.~(\ref{eq:cascadedcollision1}), using Eq.~(\ref{eq:cascadedcollisionoperator}) as
\begin{equation}
(\mathcal{K}\cdot \mathbf{\widehat{g}})_{\alpha}=(\widetilde{\overline{f}}_{\alpha}-\overline{f}_{\alpha})+S_{{\alpha}},
\label{eq:transformedcollisionoperator}
\end{equation}
and reduce it by applying the central moment operator $\braket{(e_{\alpha x}-u_x)^m(e_{\alpha y}-u_y)^n|\cdot}$ on both of its sides. Thus, we get
\begin{equation}
\sum_{\beta} \braket{(e_{\alpha x}-u_x)^m(e_{\alpha y}-u_y)^n|K_{\beta}}\widehat{g}_{\beta}=(\widetilde{\widehat{\overline{\kappa}}}_{x^my^n}-\widehat{\overline{\kappa}}_{x^my^n})+\widehat{\sigma}_{x^my^n}.
\label{eq:centralmomentcollisionstep}
\end{equation}
Let us now consider a specific case when the post-collision state is in ``equilibrium state''. In this case, we set
\begin{equation}
\widetilde{\widehat{\overline{\kappa}}}_{x^my^n}=\widehat{\overline{\kappa}}_{x^my^n}^{eq}, \widehat{\sigma}_{x^my^n}=0 \Rightarrow \widehat{g}_{\beta}=\widehat{g}_{\beta}^{*}
\label{eq:specialcasecascaded}
\end{equation}
so that $\widehat{g}_{\beta}$ takes certain specific values, $\widehat{g}_{\beta}^{*}$.

Thus the specialized non-conserved collision kernel can be obtained by expanding the LHS of Eq.~(\ref{eq:centralmomentcollisionstep}) and using Eq.~(\ref{eq:specialcasecascaded}) for $m+n \geq 2$, which can be written in matrix form as
\begin{equation}
\mathcal{F}\left[
\begin{array}{l}
{\widehat{g}_3^{*}}\\
{\widehat{g}_4^{*}}\\
{\widehat{g}_5^{*}}\\
{\widehat{g}_6^{*}}\\
{\widehat{g}_7^{*}}\\
{\widehat{g}_8^{*}}
\end{array}
\right]=\left[
\begin{array}{l}
{\widehat{\overline{\kappa}}_{xx}^{eq}-\widehat{\overline{\kappa}}_{xx}}\\
{\widehat{\overline{\kappa}}_{yy}^{eq}-\widehat{\overline{\kappa}}_{yy}}\\
{\widehat{\overline{\kappa}}_{xy}^{eq}-\widehat{\overline{\kappa}}_{xy}}\\
{\widehat{\overline{\kappa}}_{xxy}^{eq}-\widehat{\overline{\kappa}}_{xxy}}\\
{\widehat{\overline{\kappa}}_{xyy}^{eq}-\widehat{\overline{\kappa}}_{xyy}}\\
{\widehat{\overline{\kappa}}_{xxyy}^{eq}-\widehat{\overline{\kappa}}_{xxyy}}\\
\end{array}
\right],
\label{eq:specialcasecascadedrelation}
\end{equation}
where $\mathcal{F}\equiv\mathcal{F}(\overrightarrow{x},t)$ is a $6 \times 6$ local frame transformation matrix that depends on the local fluid velocity and is given by
\begin{equation}
\mathcal{F}= \left[
\begin{array}{cccccc}
 6                   & 2                  & 0           & 0       & 0       & 0    \\
 6                   & -2                 & 0           & 0       & 0       & 0    \\
 0                   & 0                  & 4           & 0       & 0       & 0    \\
-6u_y                & -2u_y              & -8u_x       & -4      & 0       & 0    \\
-6u_x                & 2u_x               & -8u_y       & 0       & -4      & 0    \\
(8+6(u_x^2+u_y^2))   & -2(u_x^2-u_y^2)    & 16u_xu_y    & 8u_y    & 8u_x    & 4    \\
\end{array} \right].
\label{eq:frametransformationmatrix}
\end{equation}
It may be noted that Eq.~(\ref{eq:frametransformationmatrix}) has entries similar to that given in Ref.~\cite{asinari08}, except for the change in signs in the third column resulting from the specific choice made for constructing $\ket{K_5}$ in the orthogonalization (Gram-Schmidt) procedure. Now substituting for the expressions in the RHS of Eq.~(\ref{eq:specialcasecascadedrelation}) and inverting it, we get $\widehat{g}_{\beta}^{*}$ in terms of the raw moments, hydrodynamic fields and force fields. It may be written as
\begin{equation}
\left[
\begin{array}{l}
{\widehat{g}_3^{*}}\\
{\widehat{g}_4^{*}}\\
{\widehat{g}_5^{*}}\\
{\widehat{g}_6^{*}}\\
{\widehat{g}_7^{*}}\\
{\widehat{g}_8^{*}}
\end{array}
\right]=
\left[
\begin{array}{l}
{ \frac{1}{18}\rho+\frac{1}{12}\rho(u_x^2+u_y^2)-\frac{1}{12}(\widehat{\overline{\kappa}}_{xx}^{'}+\widehat{\overline{\kappa}}_{yy}^{'})-\frac{1}{12}(F_xu_x+F_yu_y)  }\\
{ \frac{1}{4}\rho(u_x^2-u_y^2)-\frac{1}{4}(\widehat{\overline{\kappa}}_{xx}^{'}-\widehat{\overline{\kappa}}_{yy}^{'})-\frac{1}{4}(F_xu_x-F_yu_y)  }\\
{
\frac{1}{4}\rho u_xu_y-\frac{1}{4}\widehat{\overline{\kappa}}_{xy}^{'}-\frac{1}{8}(F_xu_y+F_yu_x)
}\\
{
-\frac{1}{12}\rho u_y-\frac{1}{4}\rho u_x^2u_y+\frac{1}{4}\widehat{\overline{\kappa}}_{xxy}^{'}+\frac{1}{4}F_xu_xu_y+\frac{1}{8}F_yu_x^2
}\\
{
-\frac{1}{12}\rho u_x-\frac{1}{4}\rho u_xu_y^2+\frac{1}{4}\widehat{\overline{\kappa}}_{xyy}^{'}+\frac{1}{4}F_yu_yu_x+\frac{1}{8}F_xu_y^2
}\\
{
-\frac{1}{12}\rho -\frac{1}{12}\rho (u_x^2+u_y^2)+\frac{1}{4}\rho u_x^2u_y^2+\frac{1}{6}(\widehat{\overline{\kappa}}_{xx}^{'}+\widehat{\overline{\kappa}}_{yy}^{'})
-\frac{1}{4}\widehat{\overline{\kappa}}_{xxyy}^{'}+q_{xxyy}
}
\end{array}
\right],
\label{eq:expressionsforspecialg}
\end{equation}
where $q_{xxyy}=\frac{1}{6}(F_xu_x+F_yu_y)-\frac{1}{4}(F_xu_xu_y^2+F_yu_yu_x^2)$. An alternative and a somewhat
direct procedure to obtain $\widehat{g}_{\beta}^{*}$ is to invoke the orthogonal properties of the basis vectors
$\ket{K_{\beta}}$. Accordingly, we can write
\begin{equation}
\widehat{g}_{\beta}^{*}=\frac{\braket{\overline{f}_{\alpha}^{eq}-\overline{f}_{\alpha}|K_{\beta}}}{\braket{K_{\beta}|K_{\beta}}}
=\frac{\braket{f_{\alpha}^{eq}-\overline{f}_{\alpha}-\frac{1}{2}S_{\alpha}|K_{\beta}}}{\braket{K_{\beta}|K_{\beta}}},
\quad \quad \beta=3,4,5,\ldots,8,
\label{eq:simplifiedcollisionkernel}
\end{equation}
which gives expressions identical to that given in Eq.~(\ref{eq:expressionsforspecialg}).

Equivalently, for the special case noted above (Eq.~(\ref{eq:specialcasecascaded})), the collision operator, Eq.~(\ref{eq:transformedcollisionoperator}), can also be written as
$\mathcal{K}\cdot\mathbf{\widehat{g}}^{*}=\mathbf{\overline{f}}^{eq}-\mathbf{\overline{f}}=\mathbf{f}^{eq}-\mathbf{\overline{f}}-\frac{1}{2}\mathbf{S}$,
which can be inverted to yield
\begin{equation}
\mathbf{\widehat{g}}^{*}=\mathcal{K}^{-1}\left(\mathbf{f}^{eq}-\mathbf{\overline{f}}-\frac{1}{2}\mathbf{S}\right),
\label{eq:specialgvector}
\end{equation}
where as before the boldface symbols represent the column vectors. Now, we propose to ``over-relax'' the above special system by means of multiple relaxation times (MRT) as a representation of collision process. That is, we set \begin{equation}
\mathbf{\widehat{g}}=\Lambda\mathbf{\widehat{g}}^{*},
\label{eq:overrelaxation}
\end{equation}
where $\Lambda$ is a relaxation time matrix. Hence, combining Eqs.~(\ref{eq:specialgvector}) and (\ref{eq:overrelaxation}), we can write the post-collision state in this MRT formulation as
\begin{eqnarray}
\mathbf{\widetilde{\overline{f}}}
=\mathbf{\overline{f}}+\mathcal{K}\cdot\mathbf{\widehat{g}}+\mathbf{S}
&=&\mathbf{\overline{f}}+\mathcal{K}\Lambda\mathbf{\widehat{g}}^{*}+\mathbf{S}\nonumber\\
&=&\mathbf{\overline{f}}+\mathcal{K}\Lambda\mathcal{K}^{-1}\left(\mathbf{f}^{eq}-\mathbf{\overline{f}}-\frac{1}{2}\mathbf{S}\right)+\mathbf{S}
\label{eq:newpostcollision}
\end{eqnarray}
Let,
\begin{equation}
\Lambda^{*}=\mathcal{K}\Lambda\mathcal{K}^{-1}.
\end{equation}
Hence,
\begin{equation}
\mathbf{\widetilde{\overline{f}}}
=\mathbf{\overline{f}}+\Lambda^{*}\left(\mathbf{f}^{eq}-\mathbf{\overline{f}}\right)+\left(\mathcal{I}-\frac{1}{2}\Lambda^{*}\right)\mathbf{S}
\label{eq:newpostcollision1}
\end{equation}
where $\mathcal{I}$ is the identity matrix.

We now define raw moments of distribution functions (including the transformed one), equilibrium and sources for convenience as
\begin{equation}
\mathbf{\widehat{\overline{f}}}=\mathcal{T}\mathbf{\overline{f}}, \quad
\mathbf{\widehat{f}}=\mathcal{T}\mathbf{f},\quad
\mathbf{\widehat{f}}^{eq}=\mathcal{T}\mathbf{f}^{eq}, \quad
\mathbf{\widehat{S}}=\mathcal{T}\mathbf{S},
\label{eq:momenttransformation}
\end{equation}
where $\widehat{(\cdot)}$ represents column vectors in (raw) moment space and the transformation matrix $\mathcal{T}$ is given in
Eq.~(\ref{eq:nominalmomentbasis}). That is,
\begin{eqnarray*}
\mathbf{\widehat{\overline{f}}}=\left(\widehat{\overline{f}}_{0},\widehat{\overline{f}}_{1},\widehat{\overline{f}}_{2},\ldots,\widehat{\overline{f}}_{8}\right)^{\dag}
&=&\left(\widehat{\overline{\kappa}}_{0}^{'},\widehat{\overline{\kappa}}_{x}^{'},\widehat{\overline{\kappa}}_{y}^{'},
\widehat{\overline{\kappa}}_{xx}^{'}+\widehat{\overline{\kappa}}_{yy}^{'},\widehat{\overline{\kappa}}_{xx}^{'}-\widehat{\overline{\kappa}}_{yy}^{'},
\widehat{\overline{\kappa}}_{xy}^{'},\widehat{\overline{\kappa}}_{xxy}^{'},\widehat{\overline{\kappa}}_{xyy}^{'},
\widehat{\overline{\kappa}}_{xxyy}^{'}\right)^{\dag},\\
\mathbf{\widehat{f}}=\left(\widehat{f}_{0},\widehat{f}_{1},\widehat{f}_{2},\ldots,\widehat{f}_{8}\right)^{\dag}
&=&\left(\widehat{\kappa}_{0}^{'},\widehat{\kappa}_{x}^{'},\widehat{\kappa}_{y}^{'},
\widehat{\kappa}_{xx}^{'}+\widehat{\kappa}_{yy}^{'},\widehat{\kappa}_{xx}^{'}-\widehat{\kappa}_{yy}^{'},
\widehat{\kappa}_{xy}^{'},\widehat{\kappa}_{xxy}^{'},\widehat{\kappa}_{xyy}^{'},
\widehat{\kappa}_{xxyy}^{'}\right)^{\dag},\\
\mathbf{\widehat{f}}^{eq}=\left(\widehat{f}_{0}^{eq},\widehat{f}_{1}^{eq},\widehat{f}_{2}^{eq},\ldots,\widehat{f}_{8}^{eq}\right)^{\dag}
&=&\left(\widehat{\kappa}_{0}^{eq'},\widehat{\kappa}_{x}^{eq'},\widehat{\kappa}_{y}^{eq'},
\widehat{\kappa}_{xx}^{eq'}+\widehat{\kappa}_{yy}^{eq'},\widehat{\kappa}_{xx}^{eq'}-\widehat{\kappa}_{yy}^{eq'},
\widehat{\kappa}_{xy}^{eq'},\widehat{\kappa}_{xxy}^{eq'},\widehat{\kappa}_{xyy}^{eq'},
\widehat{\kappa}_{xxyy}^{eq'}\right)^{\dag},\\
\mathbf{\widehat{S}}=\left(\widehat{S}_{0},\widehat{S}_{1},\widehat{S}_{2},\ldots,\widehat{S}_{8}\right)^{\dag}
&=&\left(\widehat{\sigma}_{0}^{'},\widehat{\sigma}_{x}^{'},\widehat{\sigma}_{y}^{'},
\widehat{\sigma}_{xx}^{'}+\widehat{\sigma}_{yy}^{'},\widehat{\sigma}_{xx}^{'}-\widehat{\sigma}_{yy}^{'},
\widehat{\sigma}_{xy}^{'},\widehat{\sigma}_{xxy}^{'},\widehat{\sigma}_{xyy}^{'},
\widehat{\sigma}_{xxyy}^{'}\right)^{\dag}.
\end{eqnarray*}

Finally, using Eq.~(\ref{eq:momenttransformation}), we can rewrite the expressions for the collision and source terms in Eq.~(\ref{eq:newpostcollision1}) in terms of (raw) moment space. That is,
\begin{equation}
\mathbf{\widetilde{\overline{f}}}
=\mathbf{\overline{f}}+\mathcal{T}^{-1}\left[-\widehat{\Lambda}\left(\mathbf{\widehat{\overline{f}}}-\mathbf{\widehat{f}}^{eq}\right)+\left(\mathcal{I}-\frac{1}{2}\widehat{\Lambda}\right)\mathbf{\widehat{S}}\right],
\label{eq:newpostcollision2}
\end{equation}
where $\widehat{\Lambda}$ is a diagonal collision matrix given by
\begin{equation}
\widehat{\Lambda}=\mathcal{T}\Lambda^{*}\mathcal{T}^{-1}=diag(0,0,0,\omega_3,\omega_4,\omega_5,\omega_6,\omega_7,\omega_8).
\end{equation}

It may be noted that from Eq.~(\ref{eq:momenttransformation}), we can obtain the discrete equilibrium distribution functions and source terms in velocity space by means of the inverse transformation. That is, $\mathbf{f}^{eq}=\mathcal{T}^{-1}\mathbf{\widehat{f}}^{eq}, \mathbf{S}=\mathcal{T}^{-1}\mathbf{\widehat{S}}$, which yield
\begin{eqnarray*}
f_{0}^{eq} &=& \frac{4}{9}\rho-\frac{2}{3}\rho(u_x^2+u_y^2)+\rho u_x^2u_y^2,\\
f_{1}^{eq} &=& \frac{1}{9}\rho+\frac{1}{3}\rho u_x+\frac{1}{2}\rho u_x^2-\frac{1}{6}\rho(u_x^2+u_y^2)-\frac{1}{2}\rho u_xu_y^2-\frac{1}{2}\rho u_x^2u_y^2,\\
f_{2}^{eq} &=& \frac{1}{9}\rho+\frac{1}{3}\rho u_y+\frac{1}{2}\rho u_y^2-\frac{1}{6}\rho(u_x^2+u_y^2)-\frac{1}{2}\rho u_x^2u_y-\frac{1}{2}\rho u_x^2u_y^2,\\
f_{3}^{eq} &=& \frac{1}{9}\rho-\frac{1}{3}\rho u_x+\frac{1}{2}\rho u_x^2-\frac{1}{6}\rho(u_x^2+u_y^2)+\frac{1}{2}\rho u_xu_y^2-\frac{1}{2}\rho u_x^2u_y^2,\\
f_{4}^{eq} &=& \frac{1}{9}\rho-\frac{1}{3}\rho u_y+\frac{1}{2}\rho u_y^2-\frac{1}{6}\rho(u_x^2+u_y^2)+\frac{1}{2}\rho u_x^2u_y-\frac{1}{2}\rho u_x^2u_y^2,\\
f_{5}^{eq} &=& \frac{1}{36}\rho+\frac{1}{12}\rho u_x+\frac{1}{12}\rho u_y+\frac{1}{12}\rho(u_x^2+u_y^2)+\frac{1}{4}\rho u_xu_y+\frac{1}{4}\rho u_x^2u_y+\frac{1}{4}\rho u_xu_y^2+\frac{1}{4}\rho u_x^2u_y^2,\\
f_{6}^{eq} &=& \frac{1}{36}\rho-\frac{1}{12}\rho u_x+\frac{1}{12}\rho u_y+\frac{1}{12}\rho(u_x^2+u_y^2)-\frac{1}{4}\rho u_xu_y+\frac{1}{4}\rho u_x^2u_y-\frac{1}{4}\rho u_xu_y^2+\frac{1}{4}\rho u_x^2u_y^2,\\
f_{7}^{eq} &=& \frac{1}{36}\rho-\frac{1}{12}\rho u_x-\frac{1}{12}\rho u_y+\frac{1}{12}\rho(u_x^2+u_y^2)+\frac{1}{4}\rho u_xu_y-\frac{1}{4}\rho u_x^2u_y-\frac{1}{4}\rho u_xu_y^2+\frac{1}{4}\rho u_x^2u_y^2,\\
f_{8}^{eq} &=& \frac{1}{36}\rho+\frac{1}{12}\rho u_x-\frac{1}{12}\rho u_y+\frac{1}{12}\rho(u_x^2+u_y^2)-\frac{1}{4}\rho u_xu_y-\frac{1}{4}\rho u_x^2u_y+\frac{1}{4}\rho u_xu_y^2+\frac{1}{4}\rho u_x^2u_y^2,
\end{eqnarray*}
and
\begin{eqnarray*}
S_{0}&=&-2F_xu_x-2F_yu_y+2F_xu_xu_y^2+2F_yu_yu_x^2,\\
S_{1}&=&+\frac{1}{2}F_x+F_xu_x-\frac{1}{2}F_xu_y^2-F_yu_yu_x-F_xu_xu_y^2-F_yu_yu_x^2,\\
S_{2}&=&+\frac{1}{2}F_y+F_yu_y-\frac{1}{2}F_yu_x^2-F_xu_xu_y-F_xu_xu_y^2-F_yu_yu_x^2,\\
S_{3}&=&-\frac{1}{2}F_x+F_xu_x+\frac{1}{2}F_xu_y^2+F_yu_yu_x-F_xu_xu_y^2-F_yu_yu_x^2,\\
S_{4}&=&-\frac{1}{2}F_y+F_yu_y+\frac{1}{2}F_yu_x^2+F_xu_xu_y-F_xu_xu_y^2-F_yu_yu_x^2,\\
S_{5}&=&+\frac{1}{4}F_xu_y+\frac{1}{4}F_yu_x+\frac{1}{4}F_xu_y^2+\frac{1}{4}F_yu_x^2+\frac{1}{2}F_xu_xu_y+\frac{1}{2}F_yu_yu_x+
        \frac{1}{2}F_xu_xu_y^2+\frac{1}{2}F_yu_yu_x^2,\\
S_{6}&=&-\frac{1}{4}F_xu_y-\frac{1}{4}F_yu_x-\frac{1}{4}F_xu_y^2+\frac{1}{4}F_yu_x^2+\frac{1}{2}F_xu_xu_y-\frac{1}{2}F_yu_yu_x+
        \frac{1}{2}F_xu_xu_y^2+\frac{1}{2}F_yu_yu_x^2,\\
S_{7}&=&+\frac{1}{4}F_xu_y+\frac{1}{4}F_yu_x-\frac{1}{4}F_xu_y^2-\frac{1}{4}F_yu_x^2-\frac{1}{2}F_xu_xu_y-\frac{1}{2}F_yu_yu_x+
        \frac{1}{2}F_xu_xu_y^2+\frac{1}{2}F_yu_yu_x^2,\\
S_{8}&=&-\frac{1}{4}F_xu_y-\frac{1}{4}F_yu_x+\frac{1}{4}F_xu_y^2-\frac{1}{4}F_yu_x^2-\frac{1}{2}F_xu_xu_y+\frac{1}{2}F_yu_yu_x+
        \frac{1}{2}F_xu_xu_y^2+\frac{1}{2}F_yu_yu_x^2.
\end{eqnarray*}
Thus, the discrete equilibrium distribution and forcing terms in velocity space resulting from corresponding imposed central moments are proper polynomials containing higher order terms as compared to the standard LBM. The specific functional expressions for $f_{\alpha}^{eq}$ and $S_{\alpha}$ depend on the choice made for the ``nominal moment basis'' (Eq.~(\ref{eq:nominalmomentbasis})) from which they are derived.

We are now in a position to perform a Chapman-Enskog multiscale expansion. First, expand the raw moments $\mathbf{\widehat{f}}$ (untransformed ones, i.e. without ``overbar'', for simplicity) and the time derivative in terms of a small bookkeeping perturbation parameter $\epsilon$ (which will be set to $1$ at the end of the analysis)~\cite{premnath06}:
\begin{eqnarray}
\mathbf{\widehat{f}}&=&\sum_{n=0}^{\infty}\epsilon^n \mathbf{\widehat{f}}^{(n)}, \\
\partial_t&=&\sum_{n=0}^{\infty}\epsilon^n \partial_{t_n}.
\end{eqnarray}
We use a Taylor expansion for the representation of the streaming operator, which is carried out in its natural velocity space:
\begin{equation}
\mathbf{f}(\overrightarrow{x}+\overrightarrow{e}_{\alpha}\epsilon,t+\epsilon)=\sum_{n=0}^{n}\frac{\epsilon^n}{n!}(\partial_t+\overrightarrow{e}_{\alpha}\cdot\overrightarrow{\nabla})\mathbf{f}(\overrightarrow{x},t).
\end{equation}

Substituting all the above three expansions in the LBE, with Eq.~(\ref{eq:newpostcollision2}) representing the post-collision, and equating terms of the same order of successive powers of $\epsilon$ after making use of Eq.~(\ref{eq:momenttransformation}) and rearranging, we get~\cite{premnath06}:
\begin{eqnarray}
O(\epsilon^0):\quad \mathbf{\widehat{f}}^{(0)}&=&\mathbf{\widehat{f}}^{eq},\label{eq:CEzerothorder}\\
O(\epsilon^1):\quad (\partial_{t_0}+\widehat{E}_i \partial_i)\mathbf{\widehat{f}}^{(0)}&=&-\widehat{\Lambda}\mathbf{\widehat{f}}^{(1)}+\mathbf{\widehat{S}},\label{eq:CEfirstorder}\\
O(\epsilon^2):\quad \partial_{t_1}\mathbf{\widehat{f}}^{(0)}+(\partial_{t_0}+\widehat{E}_i \partial_i)\left[ \mathcal{I}-\frac{1}{2}\widehat{\Lambda}\right]\mathbf{\widehat{f}}^{(1)}&=&-\widehat{\Lambda}\mathbf{\widehat{f}}^{(2)},\label{eq:CEsecondorder}
\end{eqnarray}
where $\widehat{E}_i=\mathcal{T}(e_{\alpha i}\mathcal{I})\mathcal{T}^{-1},\quad i \in {x,y}$. After substituting for $\mathbf{\widehat{f}}^{(0)}$, $\widehat{E}_i$ and $\mathbf{\widehat{S}}$, the first-order moment equations, i.e. Eq.~(\ref{eq:CEfirstorder}) become
\begin{equation}
\partial_{t_0}\rho+\partial_x (\rho u_x)+\partial_y (\rho u_y) = 0,
\label{eq:firstorder0}
\end{equation}
\begin{equation}
\partial_{t_0}\left(\rho u_x\right)+\partial_x \left(\frac{1}{3}\rho+\rho u_x^2\right)+\partial_y \left(\rho u_xu_y\right) = F_x,
\label{eq:firstorder1}
\end{equation}
\begin{equation}
\partial_{t_0}\left(\rho u_y\right)+\partial_x \left(\rho u_xu_y\right)+\partial_y \left(\frac{1}{3}\rho+\rho u_y^2\right) = F_y,
\label{eq:firstorder2}
\end{equation}
\begin{eqnarray}
\partial_{t_0}\left(\frac{2}{3}\rho+\rho(u_x^2+u_y^2)\right)+\partial_x \left(\frac{4}{3}\rho u_x+\rho u_xu_y^2\right)&+&\partial_y \left(\frac{4}{3}\rho u_y+\rho u_x^2u_y\right)\nonumber\\ &=&-\omega_3\widehat{f}_3^{(1)}+2F_xu_x+2F_yu_y,
\label{eq:firstorder3}
\end{eqnarray}
\begin{eqnarray}
\partial_{t_0}\left(\rho(u_x^2-u_y^2)\right)+\partial_x \left(\frac{2}{3}\rho u_x-\rho u_xu_y^2\right)&+&\partial_y \left(-\frac{2}{3}\rho u_y+\rho u_x^2u_y\right)\nonumber\\
&=&-\omega_4\widehat{f}_4^{(1)}+2F_xu_x-2F_yu_y,
\label{eq:firstorder4}
\end{eqnarray}
\begin{eqnarray}
\partial_{t_0}\left(\rho u_x u_y\right)+\partial_x \left(\frac{1}{3}\rho u_y+\rho u_x^2u_y\right)&+&\partial_y \left(\frac{1}{3}\rho u_x+\rho u_xu_y^2\right)\nonumber\\
 &=&-\omega_5\widehat{f}_5^{(1)}+F_xu_y+F_yu_x,
\label{eq:firstorder5}
\end{eqnarray}
\begin{eqnarray}
\partial_{t_0}\left(\frac{1}{3}\rho u_y+\rho u_x^2u_y\right)+\partial_x \left(\rho u_xu_y\right)&+&\partial_y \left(\frac{1}{9}\rho+\frac{1}{3}\rho (u_x^2+u_y^2) +\rho u_x^2u_y\right)\nonumber\\ &=&-\omega_6\widehat{f}_6^{(1)}+F_yu_x^2+2F_xu_xu_y,
\label{eq:firstorder6}
\end{eqnarray}
\begin{eqnarray}
\partial_{t_0}\left(\frac{1}{3}\rho u_x+\rho u_xu_y^2\right)&+&\partial_x \left(\frac{1}{9}\rho+\frac{1}{3}\rho (u_x^2+u_y^2)+\rho u_x^2u_y^2\right)+\partial_y \left(\rho u_xu_y\right) \nonumber\\
&=&-\omega_7\widehat{f}_7^{(1)}+F_xu_y^2+2F_yu_yu_x,
\label{eq:firstorder7}
\end{eqnarray}
\begin{eqnarray}
\partial_{t_0}\left(\frac{1}{9}\rho+\frac{1}{3}\rho (u_x^2+u_y^2)+\rho u_x^2u_y^2\right)&+&\partial_x \left(\frac{1}{3}\rho u_x+\rho u_xu_y^2\right)+\partial_y \left(\frac{1}{3}\rho u_y+\rho u_x^2u_y\right) \nonumber\\ &=&-\omega_8\widehat{f}_8^{(1)}+2F_xu_xu_y^2+2F_yu_yu_xu_x^2.
\label{eq:firstorder8}
\end{eqnarray}

Similarly, the second-order moment equations can be derived from Eq.~(\ref{eq:CEsecondorder}), which can be written as
\begin{equation}
\partial_{t_0}\rho=0,
\label{eq:secondorder0}
\end{equation}
\begin{equation}
\partial_{t_1}\left(\rho u_x\right)+\partial_x \left[\frac{1}{2}\left(1-\frac{1}{2}\omega_3\right)\widehat{f}_3^{(1)}+\frac{1}{2}\left(1-\frac{1}{2}\omega_4\right)\widehat{f}_4^{(1)}\right]
+\partial_y \left[\left(1-\frac{1}{2}\omega_5\right)\widehat{f}_5^{(1)}\right] =0,
\label{eq:secondorder1}
\end{equation}
\begin{equation}
\partial_{t_1}\left(\rho u_y\right)+\partial_x \left[\left(1-\frac{1}{2}\omega_5\right)\widehat{f}_5^{(1)}\right]+\partial_y \left[\frac{1}{2}\left(1-\frac{1}{2}\omega_3\right)\widehat{f}_3^{(1)}-\frac{1}{2}\left(1-\frac{1}{2}\omega_4\right)\widehat{f}_4^{(1)}\right]
=0,
\label{eq:secondorder2}
\end{equation}
\begin{eqnarray}
\partial_{t_1}\left(\frac{2}{3}\rho+\rho(u_x^2+u_y^2)\right)
&+&\partial_{t_0}\left[\left(1-\frac{1}{2}\omega_3\right)\widehat{f}_3^{(1)}\right]
+\partial_x \left[\left(1-\frac{1}{2}\omega_7\right)\widehat{f}_7^{(1)}\right] \nonumber\\
&+&\partial_y \left[\left(1-\frac{1}{2}\omega_6\right)\widehat{f}_6^{(1)}\right]= -\omega_3 \widehat{f}_3^{(2)},
\label{eq:secondorder3}
\end{eqnarray}
\begin{eqnarray}
\partial_{t_1}\left(\rho(u_x^2-u_y^2)\right)
&+&\partial_{t_0}\left[\left(1-\frac{1}{2}\omega_4\right)\widehat{f}_4^{(1)}\right]
+\partial_x \left[-\left(1-\frac{1}{2}\omega_7\right)\widehat{f}_7^{(1)}\right] \nonumber\\
&+&\partial_y \left[\left(1-\frac{1}{2}\omega_6\right)\widehat{f}_6^{(1)}\right]= -\omega_4 \widehat{f}_4^{(2)},
\label{eq:secondorder4}
\end{eqnarray}
\begin{eqnarray}
\partial_{t_1}\left(\rho u_xu_y\right)
&+&\partial_{t_0}\left[\left(1-\frac{1}{2}\omega_5\right)\widehat{f}_5^{(1)}\right]
+\partial_x \left[\left(1-\frac{1}{2}\omega_6\right)\widehat{f}_6^{(1)}\right] \nonumber\\
&+&\partial_y \left[\left(1-\frac{1}{2}\omega_7\right)\widehat{f}_7^{(1)}\right]= -\omega_5 \widehat{f}_5^{(2)},
\label{eq:secondorder5}
\end{eqnarray}
\begin{eqnarray}
\partial_{t_1}\left(\frac{1}{3}\rho u_y+\rho u_x^2u_y\right)
&+&\partial_{t_0}\left[\left(1-\frac{1}{2}\omega_6\right)\widehat{f}_6^{(1)}\right]
+\partial_x \left[\left(1-\frac{1}{2}\omega_5\right)\widehat{f}_5^{(1)}\right] \nonumber\\
&+&\partial_y \left[\left(1-\frac{1}{2}\omega_8\right)\widehat{f}_8^{(1)}\right]= -\omega_6 \widehat{f}_6^{(2)},
\label{eq:secondorder6}
\end{eqnarray}
\begin{eqnarray}
\partial_{t_1}\left(\frac{1}{3}\rho u_x+\rho u_xu_y^2\right)
&+&\partial_{t_0}\left[\left(1-\frac{1}{2}\omega_7\right)\widehat{f}_7^{(1)}\right]
+\partial_x \left[\left(1-\frac{1}{2}\omega_8\right)\widehat{f}_8^{(1)}\right] \nonumber\\
&+&\partial_y \left[\left(1-\frac{1}{2}\omega_5\right)\widehat{f}_5^{(1)}\right]= -\omega_7 \widehat{f}_7^{(2)},
\label{eq:secondorder7}
\end{eqnarray}
\begin{eqnarray}
\partial_{t_1}\left(\frac{1}{9}\rho+\frac{1}{3}\rho (u_x^2+u_y^2)+\rho u_x^2u_y^2 \right)
&+&\partial_{t_0}\left[\left(1-\frac{1}{2}\omega_8\right)\widehat{f}_8^{(1)}\right]
+\partial_x \left[\left(1-\frac{1}{2}\omega_7\right)\widehat{f}_7^{(1)}\right] \nonumber\\
&+&\partial_y \left[\left(1-\frac{1}{2}\omega_6\right)\widehat{f}_6^{(1)}\right]= -\omega_8 \widehat{f}_8^{(2)}.
\label{eq:secondorder8}
\end{eqnarray}

Combining Eqs.~(\ref{eq:firstorder0}), (\ref{eq:firstorder1}) and (\ref{eq:firstorder2}), with $\epsilon$ times
Eqs.~(\ref{eq:secondorder0}), (\ref{eq:secondorder1}) and (\ref{eq:secondorder2}), respectively, and using
$\partial_t=\partial_{t_0}+\epsilon \partial_{t_1}$, we get the dynamical equations for the conserved or hydrodynamic
moments after setting the parameter $\epsilon$ to unity. That is,
\begin{equation}
\partial_t \rho +\partial_x (\rho u_x)+ \partial_y (\rho u_y) = 0,
\label{eq:hydromoments0}
\end{equation}
\begin{eqnarray}
\partial_t (\rho u_x) +\partial_x (\rho u_x^2)+ \partial_y (\rho u_xu_y) &=& - \partial_x\left(\frac{1}{3}\rho\right)
-\partial_x \left[ \frac{1}{2}\left(1-\frac{1}{2}\omega_3\right)\widehat{f}_3^{(1)}+\frac{1}{2}\left(1-\frac{1}{2}\omega_4\right)\widehat{f}_4^{(1)} \right] \nonumber \\
&& -\partial_y \left[ \left(1-\frac{1}{2}\omega_5\right)\widehat{f}_5^{(1)} \right]+F_x,
\label{eq:hydromoments1}
\end{eqnarray}
\begin{eqnarray}
\partial_t (\rho u_y) +\partial_x (\rho u_x u_y)&+&\partial_y (\rho u_y^2) = - \partial_x\left(\frac{1}{3}\rho\right)
-\partial_x \left[ \left(1-\frac{1}{2}\omega_5\right)\widehat{f}_5^{(1)} \right]\nonumber \\
&-&\partial_y\left[ \frac{1}{2}\left(1-\frac{1}{2}\omega_3\right)\widehat{f}_3^{(1)}-\frac{1}{2}\left(1-\frac{1}{2}\omega_4\right)\widehat{f}_4^{(1)} \right]+F_y.
\label{eq:hydromoments2}
\end{eqnarray}
In the above three equations, Eqs.~(\ref{eq:hydromoments0})-(\ref{eq:hydromoments2}), we need the non-equilibrium raw moments
$\widehat{f}_3^{(1)}$, $\widehat{f}_4^{(1)}$ and $\widehat{f}_5^{(1)}$ or $\widehat{\pi}_{xx}^{'(1)}+\widehat{\pi}_{yy}^{'(1)}$, $\widehat{\pi}_{xx}^{'(1)}-\widehat{\pi}_{yy}^{'(1)}$ and $\widehat{\pi}_{xy}^{'(1)}$, respectively. They can be obtained from Eqs.~(\ref{eq:firstorder4}), (\ref{eq:firstorder5}) and (\ref{eq:firstorder6}), respectively. Thus,
\begin{eqnarray}
\widehat{f}_3^{(1)} &=& \frac{1}{\omega_3}\left[\left\{ -\partial_{t_0}\left(\frac{2}{3}\rho+\rho (u_x^2+u_y^2)\right)
-\partial_x\left(\frac{4}{3}\rho u_x+\rho u_xu_y^2\right)
-\partial_y\left(\frac{4}{3}\rho u_y+\rho u_x^2u_y\right)
 \right\}\right.\nonumber\\
 &&\left.+2F_xu_x+2F_yu_y\right],
\label{eq:noneqmfhat3}
\end{eqnarray}
\begin{eqnarray}
\widehat{f}_4^{(1)} &=& \frac{1}{\omega_4}\left[\left\{ -\partial_{t_0}\left(\rho (u_x^2-u_y^2)\right)
-\partial_x\left(\frac{2}{3}\rho u_x-\rho u_xu_y^2\right)
-\partial_y\left(-\frac{2}{3}\rho u_y+\rho u_x^2u_y\right)
 \right\}\right.\nonumber\\
 &&\left.+2F_xu_x-2F_yu_y\right],
\label{eq:noneqmfhat4}
\end{eqnarray}
\begin{eqnarray}
\widehat{f}_5^{(1)} &=& \frac{1}{\omega_5}\left[\left\{ -\partial_{t_0}\left(\rho u_x u_y\right)
-\partial_x\left(\frac{1}{3}\rho u_y+\rho u_x^2u_y\right)
-\partial_y\left(\frac{1}{3}\rho u_x+\rho u_xu_y^2\right)
 \right\}\right.\nonumber\\
 &&\left.+F_xu_y+F_yu_x\right],
\label{eq:noneqmfhat5}
\end{eqnarray}
The above three non-equilibrium moments can be simplified. In particular, by using the first-order hydrodynamic moment equations, Eqs.~(\ref{eq:firstorder0})-(\ref{eq:firstorder2}) and neglecting terms of $O(u^3)$ or higher, we have
$\partial_{t_0}(\rho u_x^2)\approx 2F_xu_x$, $\partial_{t_0}(\rho u_y^2)\approx 2F_yu_y$ and
$\partial_{t_0}(\rho u_xu_y)\approx F_xu_y+F_yu_x$. Substituting for these terms in Eqs.~(\ref{eq:noneqmfhat3})-(\ref{eq:noneqmfhat5}), and representing the components of momentum for brevity as
\begin{equation*}
j_x=\rho u_x, \quad j_y = \rho u_y,
\end{equation*}
we get
\begin{eqnarray}
\widehat{f}_3^{(1)} &\approx& -\frac{2}{3\omega_3}\overrightarrow{\nabla}\cdot \overrightarrow{j},\label{eq:noneqmfhat3new}\\
\widehat{f}_4^{(1)} &\approx& -\frac{2}{3\omega_4}\left[\partial_x j_x-\partial_y j_y\right],\label{eq:noneqmfhat4new}\\
\widehat{f}_5^{(1)} &\approx& -\frac{1}{3\omega_5}\left[\partial_x j_y+\partial_y j_x\right].\label{eq:noneqmfhat5new}
\end{eqnarray}

Now, let
\begin{equation}
\vartheta_3=\frac{1}{3}\left(\frac{1}{\omega_3}-\frac{1}{2}\right),\quad
\vartheta_4=\frac{1}{3}\left(\frac{1}{\omega_4}-\frac{1}{2}\right),\quad
\vartheta_5=\frac{1}{3}\left(\frac{1}{\omega_5}-\frac{1}{2}\right), \label{eq:coeffrelaxparameters}
\end{equation}
and substituting the simplified expressions for the non-conserved moments, Eqs.~(\ref{eq:noneqmfhat3new})-(\ref{eq:noneqmfhat5new}), and by using the relations for relaxation parameters given in Eq.~(\ref{eq:coeffrelaxparameters}) in the conserved moment equations, Eqs.~(\ref{eq:hydromoments0})-(\ref{eq:hydromoments2}), we get
\begin{equation}
\partial_t \rho + \overrightarrow{\nabla}\cdot \overrightarrow{j} = 0
\label{eq:hydromoments0new},
\end{equation}
\begin{eqnarray}
\partial_t j_x+\partial_x \left(\frac{j_x^2}{\rho}\right)+\partial_y \left(\frac{j_xj_y}{\rho}\right)&=&-\partial_x p+\partial_x\left[\vartheta_4(2\partial_x j_x-\overrightarrow{\nabla}\cdot\overrightarrow{j})+\vartheta_3\overrightarrow{\nabla}\cdot\overrightarrow{j}\right]\nonumber \\
&&+\partial_y\left[\vartheta_5(\partial_x j_y+\partial_y j_x) \right]+F_x,
\label{eq:hydromoments1new}
\end{eqnarray}
\begin{eqnarray}
\partial_t j_y+\partial_x \left(\frac{j_xj_y}{\rho}\right)+\partial_y \left(\frac{j_y^2}{\rho}\right)&=&-\partial_y p+\partial_x\left[\vartheta_5(\partial_x j_y+\partial_y j_x)\right]\nonumber \\
&&+\partial_y\left[\vartheta_4(2\partial_y j_y-\overrightarrow{\nabla}\cdot\overrightarrow{j})+\vartheta_3\overrightarrow{\nabla}\cdot\overrightarrow{j} \right]+F_y,
\label{eq:hydromoments2new}
\end{eqnarray}
where $p=\frac{1}{3}\rho$ is the pressure field. Evidently, the relaxation parameters $\omega_4$ and $\omega_5$ determine the
shear kinematic viscosity of the fluid, while $\omega_3$ controls its bulk viscous behavior. Moreover, $\omega_4=\omega_5$ to maintain isotropy of the viscous stress tensor ($\vartheta_4=\vartheta_5$). Thus, the proposed semi-implicit procedure for incorporating forcing term based on a specialized central moment lattice kinetic formulation is consistent with the weakly compressible Navier-Stokes equations without resulting in any spurious effects.

It may be noted that in this work, we have employed a multiscale, or more specifically a two time scale, expansion~\cite{chapman64} to derive the macroscopic equations. An alternative approach is to consider a single time scale with an appropriate scaling relationship between space step and time step to recover specific type of fluid flow behavior. This broadly leads to two different types of consistency analysis techniques: (a) asymptotic analysis approach~\cite{sone02} based on a diffusive or parabolic scaling~\cite{junk05} and (b) equivalent equation approach used in conjunction with certain smoothness assumption and Taylor series expansion~\cite{lerat74,warming74} based on a convective or hyperbolic scaling~\cite{dubois08}. A recursive application of the LBE and an associated Taylor series expansion without an explicit asymptotic relationship between the lattice parameters can also be used to analyze the structure of the truncation errors of the emergent macroscopic equations~\cite{holdych04}. Another more recently developed approach is that based on a truncated Grad moment expansion using appropriate scaling with a recursive substitution procedure~\cite{asinari08}, which has some features in common with an order of magnitude analysis for kinetic methods~\cite{struchtrup05}. It is expected that such analysis tools can alternatively be applied to study the new computational approach described in this work.

\section{\label{app:GeneralizedEqilibriumSources}Generalization of Equilibrium and Sources with a Multiple Relaxation Time Cascaded Lattice Kinetic Formulation}
Let us first consider relaxation process of second-order non-conserved moments in the rest frame of reference:
\begin{equation}
\widehat{g}_{\beta}^{c}=\omega_{\beta}g_{\beta}^{*}= \omega_{\beta}\frac{\braket{\overline{f}_{\alpha}^{eq}-\overline{f}_{\alpha}|K_{\beta}}}{\braket{K_{\beta}|K_{\beta}}},\quad \beta=3,4,5.
\end{equation}
Here, summation of repeated indices with the subscript $\beta$ on the RHS is not assumed and the superscript ``c'' for $\widehat{g}_{\beta}$ represents its evaluation for cascaded collision process, with $g_{\beta}^{*}$ given in Eq.~(\ref{eq:simplifiedcollisionkernel}) but restrict here to second-order moments. For convenience, we now define the non-equilibrium (raw) moment of order $(m+n)$ as
\begin{equation}
\widehat{\overline{\kappa}}_{x^my^n}^{(neq)'}=\widehat{\overline{\kappa}}_{x^my^n}^{'}-\widehat{\overline{\kappa}}_{x^my^n}^{eq'},
\label{eq:noneqmrawmoment}
\end{equation}
or equivalently $\widehat{\overline{f}}_{\beta}^{(neq)}=\widehat{\overline{f}}_{\beta}-\widehat{\overline{f}}_{\beta}^{eq}$, where $\beta=m+n$. Thus,
\begin{eqnarray}
\widehat{g}_{3}^{c}&=&-\frac{\omega_3}{12}\left[\widehat{\overline{\kappa}}_{xx}^{(neq)'}+\widehat{\overline{\kappa}}_{yy}^{(neq)'}\right]=-\frac{\omega_3}{12}\widehat{\overline{f}}_{3}^{(neq)},\\
\widehat{g}_{4}^{c}&=&-\frac{\omega_4}{4}\left[\widehat{\overline{\kappa}}_{xx}^{(neq)'}-\widehat{\overline{\kappa}}_{yy}^{(neq)'}\right]=-\frac{\omega_4}{4}\widehat{\overline{f}}_{4}^{(neq)},\\
\widehat{g}_{5}^{c}&=&-\frac{\omega_5}{4}\left[\widehat{\overline{\kappa}}_{xy}^{(neq)'}\right]=-\frac{\omega_5}{4}\widehat{\overline{f}}_{5}^{(neq)},
\end{eqnarray}

The next step is to relax the third and higher order non-conserved moments in the moving frame of reference, with each \emph{central} moment relaxing with distinct relaxation time, in general. That is,
\begin{equation}
\sum_{\beta} \braket{(e_{\alpha x}-u_x)^m(e_{\alpha y}-u_y)^n|K_{\beta}}\widehat{g}_{\beta}^{c}=\omega_{\beta}\left[\widehat{\overline{\kappa}}_{x^my^n}^{eq}-\widehat{\overline{\kappa}}_{x^my^n}+\widehat{\sigma}_{x^my^n}\right], \quad m+n\geq 3.
\label{eq:highercentralmomentrelaxation}
\end{equation}
Clearly, this is equivalent to considering the last three rows of the $\mathcal{F}$ matrix given in Eq.~(\ref{eq:frametransformationmatrix}) to determine $\widehat{g}_{\beta}^{c}$, for $\beta=6,7,8$~\cite{asinari08}. Expanding the
terms within the brackets of the RHS Eq.~(\ref{eq:highercentralmomentrelaxation}) in terms of raw moments, we get
\begin{eqnarray}
\widehat{\overline{\kappa}}_{xxy}^{eq}-\widehat{\overline{\kappa}}_{xxy}-\widehat{\sigma}_{xxy}&=&-\left[\widehat{\overline{\kappa}}_{xxy}^{(neq)'}-2u_x\widehat{\overline{\kappa}}_{xy}^{(neq)'}-u_y\widehat{\overline{\kappa}}_{xx}^{(neq)')}\right],\label{eq:relationbetweencentralandrawxxy}\\
\widehat{\overline{\kappa}}_{xyy}^{eq}-\widehat{\overline{\kappa}}_{xyy}-\widehat{\sigma}_{xyy}&=&-\left[\widehat{\overline{\kappa}}_{xyy}^{(neq)'}-2u_y\widehat{\overline{\kappa}}_{xy}^{(neq)'}-u_x\widehat{\overline{\kappa}}_{yy}^{(neq)')}\right],\label{eq:relationbetweencentralandrawxyy}\\
\widehat{\overline{\kappa}}_{xxyy}^{eq}-\widehat{\overline{\kappa}}_{xxyy}-\widehat{\sigma}_{xxyy}&=&-\left[\widehat{\overline{\kappa}}_{xxyy}^{(neq)'}-2u_x\widehat{\overline{\kappa}}_{xyy}^{(neq)'}-2u_y\widehat{\overline{\kappa}}_{xxy}^{(neq)')}+u_x^2\widehat{\overline{\kappa}}_{yy}^{(neq)')}+u_y^2\widehat{\overline{\kappa}}_{xx}^{(neq)')}\right.\nonumber\\&&\left.+4u_xu_y\widehat{\overline{\kappa}}_{xy}^{(neq)')}\right].\label{eq:relationbetweencentralandrawxxyy}
\end{eqnarray}

Now, in a manner analogous to the relaxation of second-order (raw) moments to their equilibrium states, we assume relaxation of third and higher order (raw) moments to their corresponding ``equilibrium'' states as well, which are as yet unknown, but will be determined in the following consideration. That is, we consider the ansatz
\begin{equation}
\widehat{g}_{\beta}^{c}= \omega_{\beta}\frac{\braket{\overline{f}_{\alpha}^{eq,G}-\overline{f}_{\alpha}|K_{\beta}}}{\braket{K_{\beta}|K_{\beta}}},\quad \beta=6,7,8.
\label{eq:generalizedcascadedcollision}
\end{equation}
Here, the superscript ``G'' represents the ``generalized'' expression, i.e. $\overline{f}_{\alpha}^{eq,G}$ is the generalized equilibrium in the presence of forcing terms (due to the presence of the `overbar' symbol), which for $\alpha=6,7,8$ will be determined in the following.  Again, summation of repeated indices with the subscript $\beta$ on the RHS is not assumed. Evaluating Eq.~(\ref{eq:generalizedcascadedcollision}) yields
\begin{eqnarray}
\widehat{g}_{6}^{c}&=&-\frac{\omega_6}{4}\left[\widehat{\overline{\kappa}}_{xxy}^{eq,G'}-\widehat{\overline{\kappa}}_{xxy}^{'}\right],\\
\widehat{g}_{7}^{c}&=&-\frac{\omega_7}{4}\left[\widehat{\overline{\kappa}}_{xyy}^{eq,G'}-\widehat{\overline{\kappa}}_{xyy}^{'}\right],\\
\widehat{g}_{8}^{c}&=&\frac{\omega_8}{4}\left[\widehat{\overline{\kappa}}_{xxyy}^{eq,G'}-\widehat{\overline{\kappa}}_{xxyy}^{'}\right]
-\frac{\omega_8}{4}\left[\widehat{\overline{\kappa}}_{xx}^{eq'}-\widehat{\overline{\kappa}}_{xx}^{'}\right]
-\frac{\omega_8}{4}\left[\widehat{\overline{\kappa}}_{yy}^{eq'}-\widehat{\overline{\kappa}}_{yy}^{'}\right].
\end{eqnarray}
Now substituting Eqs.~(\ref{eq:noneqmrawmoment}),(\ref{eq:relationbetweencentralandrawxxy})-(\ref{eq:relationbetweencentralandrawxxyy}) and (\ref{eq:generalizedcascadedcollision}) in Eq.~(\ref{eq:highercentralmomentrelaxation}) and simplifying and rearranging the resulting expressions yield the desired expressions for the generalized equilibrium in the presence of forcing terms
\begin{eqnarray}
\widehat{\overline{\kappa}}_{xxy}^{eq,G'}&=&\widehat{\overline{\kappa}}_{xxy}^{eq'}+\varphi_{6}^{3}\left[\widehat{\overline{\kappa}}_{xx}^{(neq)'}
+\widehat{\overline{\kappa}}_{yy}^{(neq)'}\right]+\varphi_{6}^{4}\left[\widehat{\overline{\kappa}}_{xx}^{(neq)'}-\widehat{\overline{\kappa}}_{yy}^{(neq)'}\right]
+\varphi_{6}^{5}\widehat{\overline{\kappa}}_{xy}^{(neq)'},\label{eq:generalizedeqmxxy}\\
\widehat{\overline{\kappa}}_{xyy}^{eq,G'}&=&\widehat{\overline{\kappa}}_{xyy}^{eq'}+\varphi_{7}^{3}\left[\widehat{\overline{\kappa}}_{xx}^{(neq)'}
+\widehat{\overline{\kappa}}_{yy}^{(neq)'}\right]+\varphi_{7}^{4}\left[\widehat{\overline{\kappa}}_{xx}^{(neq)'}-\widehat{\overline{\kappa}}_{yy}^{(neq)'}\right]
+\varphi_{7}^{5}\widehat{\overline{\kappa}}_{xy}^{(neq)'},\label{eq:generalizedeqmxyy}\\
\widehat{\overline{\kappa}}_{xxyy}^{eq,G'}&=&\widehat{\overline{\kappa}}_{xxyy}^{eq'}+\varphi_{8}^{3}\left[\widehat{\overline{\kappa}}_{xx}^{(neq)'}
+\widehat{\overline{\kappa}}_{yy}^{(neq)'}\right]+\varphi_{8}^{4}\left[\widehat{\overline{\kappa}}_{xx}^{(neq)'}-\widehat{\overline{\kappa}}_{yy}^{(neq)'}\right]
+\varphi_{8}^{5}\widehat{\overline{\kappa}}_{xy}^{(neq)'}\nonumber\\
&&+\varphi_{8}^{6}\widehat{\overline{\kappa}}_{xxy}^{(neq)'}+\varphi_{8}^{7}\widehat{\overline{\kappa}}_{xyy}^{(neq)'},\label{eq:generalizedeqmxxyy}
\end{eqnarray}
where the coefficients $\varphi_{\alpha}^{\beta}$ in Eqs.~(\ref{eq:generalizedeqmxxy})-(\ref{eq:generalizedeqmxxyy}) are functions of the various ratios of the relaxation times of the above MRT cascaded formalism and velocity field arising relaxing the moments in the moving frame of reference.
The coefficients for $\widehat{\overline{\kappa}}_{xxy}^{eq,G'}$ are
\begin{equation}
\varphi_{6}^{3}=\frac{1}{2}\left(1-\theta_{6}^{3}\right)u_y, \quad \varphi_{6}^{4}=\frac{1}{2}\left(1-\theta_{6}^{4}\right)u_y, \quad
\varphi_{6}^{5}=2\left(1-\theta_{6}^{5}\right)u_x,
\label{eq:coeffsgeneralizedeqm6}
\end{equation}
and for $\widehat{\overline{\kappa}}_{xyy}^{eq,G'}$ are
\begin{equation}
\varphi_{7}^{3}=\frac{1}{2}\left(1-\theta_{7}^{3}\right)u_x, \quad \varphi_{7}^{4}=-\frac{1}{2}\left(1-\theta_{7}^{4}\right)u_x, \quad
\varphi_{7}^{5}=2\left(1-\theta_{7}^{5}\right)u_y,
\label{eq:coeffsgeneralizedeqm7}
\end{equation}
and, finally, for $\widehat{\overline{\kappa}}_{xxyy}^{eq,G'}$ are
\begin{eqnarray}
\varphi_{8}^{3}&=&-\left\{\left(1-\theta_{8}^{3}\right)\left[\frac{2}{3}+\frac{1}{2}(u_x^2+u_y^2)\right]-\theta_{8}^{6}\left(1-\theta_{6}^{3}\right)u_y^2-\theta_{8}^{7}\left(1-\theta_{7}^{3}\right)u_x^2\right\},\nonumber\\
\varphi_{8}^{4}&=&\frac{1}{2}\left(1-\theta_{8}^{4}\right)(u_x^2-u_y^2)+\theta_{8}^{6}\left(1-\theta_{6}^{4}\right)u_y^2-\theta_{8}^{7}\left(1-\theta_{7}^{4}\right)u_x^2,\nonumber\\
\varphi_{8}^{5}&=&-4\left[\left(1-\theta_{8}^{5}\right)-\theta_{8}^{6}\left(1-\theta_{6}^{5}\right)-\theta_{8}^{7}\left(1-\theta_{7}^{5}\right)\right]u_xu_y,\label{eq:coeffsgeneralizedeqm8}\\
\varphi_{8}^{6}&=&2\left(1-\theta_{8}^{6}\right)u_y,\nonumber\\
\varphi_{8}^{7}&=&2\left(1-\theta_{8}^{7}\right)u_x.\nonumber
\end{eqnarray}
Here, in Eqs.~(\ref{eq:coeffsgeneralizedeqm6})-(\ref{eq:coeffsgeneralizedeqm8}), the parameter $\theta_{\beta}^{\alpha}$ refers to the ratio of relaxation times $\omega_{\alpha}$ and $\omega_{\beta}$. That is
\begin{equation}
\theta_{\beta}^{\alpha}=\frac{\omega_{\alpha}}{\omega_{\beta}}.
\end{equation}

Now, in the notations of the previous section, we can rewrite $\widehat{\overline{\kappa}}_{x^my^n}^{eq,G'}$ in terms of $\widehat{\overline{f}}_{\beta}^{G}$, or more explicitly, in terms of the regular generalized equilibrium and source moments, i.e.
$\widehat{f}_{\beta}^{G}$ and $\widehat{S}_{\beta}^{G}$, respectively, using
$\widehat{\overline{f}}_{\beta}^{G}=\widehat{f}_{\beta}^{G}-\frac{1}{2}\widehat{S}_{\beta}^{G}$. Thus, compactly, the generalized equilibrium and source moments are
\begin{eqnarray}
\widehat{f}_{\beta}^{eq,G}&=&\widehat{f}_{\beta}^{eq}+\sum_{\alpha=3}^{N_v}\varphi_{\beta}^{\alpha}\widehat{f}_{\alpha}^{(neq)}=\widehat{f}_{\beta}^{eq}+\sum_{\alpha=3}^{N_v}\varphi_{\beta}^{\alpha}\left(\widehat{f}_{\alpha}-\widehat{f}_{\alpha}^{(eq)}\right), \quad \quad \beta=6,7,8 \label{eq:generalizedequilibrium}\\
\widehat{S}_{\beta}^{G}&=&\widehat{S}_{\beta}-\sum_{\alpha=3}^{N_v}\varphi_{\beta}^{\alpha}\widehat{S}_{\alpha}, \quad \quad \beta=6,7,8 \label{eq:generalizedsource}
\end{eqnarray}
where $N_v = \left\{\begin{array}{ll}
   {5,}&{\beta=6,7}\\
   {7,}&{\beta=8}
    \end{array} \right.$. It should, however, be noted that $\widehat{f}_{\beta}^{eq,G}=\widehat{f}_{\beta}^{eq}$ and
$\widehat{S}_{\beta}^{G}=\widehat{S}_{\beta}$ for $\beta \leq 5$. This analysis further extends that of Asinari~\cite{asinari08},
who showed generalized equilibrium for a particular form of Cascaded-LBM without forcing terms. Thus, the generalized equilibrium arising from the cascaded nature of the collision step for the third and higher order (raw) moments is a function of conserved moments, non-equilibrium part of the lower order moments and the various ratios of the relaxation times in the MRT formulation. Similarly, the generalized sources for the third and higher order moments is a function of the products of force fields and fluid velocity, as well as the ratio of relaxation
times. In view of the above, the cascaded formulation can also be reinterpreted by defining the generalization of the equilibrium and source in terms of the following local coefficient matrix
$\mathcal{C}\equiv\mathcal{C}(\overrightarrow{x},t)$:
\begin{equation}
\mathcal{C}= \left[
\begin{array}{ccccccccc}
 0      & 0      & 0     & 0                 & 0                 & 0                & 0                & 0                & 0    \\
 0      & 0      & 0     & 0                 & 0                 & 0                & 0                & 0                & 0    \\
 0      & 0      & 0     & 0                 & 0                 & 0                & 0                & 0                & 0    \\
 0      & 0      & 0     & 0                 & 0                 & 0                & 0                & 0                & 0    \\
 0      & 0      & 0     & 0                 & 0                 & 0                & 0                & 0                & 0    \\
 0      & 0      & 0     & 0                 & 0                 & 0                & 0                & 0                & 0    \\
 0      & 0      & 0     & \varphi_6^3       & \varphi_6^4       & \varphi_6^5      & 0                & 0                & 0    \\    0      & 0      & 0     & \varphi_7^3       & \varphi_7^4       & \varphi_7^5      & 0                & 0                & 0    \\
 0      & 0      & 0     & \varphi_8^3       & \varphi_8^4       & \varphi_8^5      & \varphi_8^6      & \varphi_8^7      & 0    \\
\end{array} \right].
\end{equation}
That is, if the information cascades from lower to higher moments during a time interval $(t,t+1)$, the raw equilibrium and source moments in the lattice frame of reference generalize to
\begin{eqnarray}
\widehat{\mathbf{f}}_{(\overrightarrow{x},t^{*})}^{eq,G}&=&\left(\mathcal{I}-\mathcal{C}\right)\widehat{\mathbf{f}}_{(\overrightarrow{x},t)}^{eq}+\mathcal{C}\widehat{\mathbf{f}}_{(\overrightarrow{x},t+1)},\\
\widehat{\mathbf{S}}_{(\overrightarrow{x},t^{*})}^{G}&=&\left(\mathcal{I}-\mathcal{C}\right)\widehat{\mathbf{S}}_{(\overrightarrow{x},t)}
\end{eqnarray}
where $t^{*}$ represents some intermediate time in $(t,t+1)$. Clearly, the generalization of both equilibrium and sources degenerate to corresponding regular forms only when the
relaxation times of all the moments are the same. That is, when the approach is reduced to the SRT formulation,
$\widehat{f}_{\beta}^{eq,G}=\widehat{f}_{\beta}^{eq}$ and $\widehat{S}_{\beta}^{G}=\widehat{S}_{\beta}$ for all possible
values of $\beta$, since $\mathcal{C}=\bf{0}$, i.e. a null matrix in that case.
In the previous section, a consistency analysis for a special case of the central moment method was presented. The same notation and procedure can be adopted for the general case involving cascaded relaxation (represented as a relaxation of non-conserved raw moments to their generalized equilibrium) with generalized sources presented here, when $\widehat{f}_{\beta}^{eq}$
becomes $\widehat{f}_{\beta}^{eq,G}$ and $\left(1-\frac{1}{2}\omega_{\beta}\right)\widehat{S}_{\beta}$ becomes $\widehat{S}_{\beta}^{G}$ for $\beta=6,7,8$. Inspection of the details of the Chapman-Enskog moment expansion analysis presented in the earlier
section shows that the consistency of the Cascaded-LBM to the NSE remains unaffected by the presence of generalized equilibrium
and sources. In particular, the generalized forms contain coefficients which are functions of local fluid velocity and the ratio of various relaxation times, and terms that are non-equilibrium part of the lower order moments, which are negligibly small in nature for slow or weakly compressible flows, as they involve products of various powers of hydrodynamic fields. Since for consistency purpose, we need to retain only $O(Ma^2)$, the presence of the generalized terms do not affect the end result of the
derivation presented in the previous section.

An interesting viewpoint to note is that the use of relaxation to generalized equilibrium (including the effect of sources), i.e. Eq.~(\ref{eq:generalizedcascadedcollision}) may be considered as an alternative computational framework to actually
execute the cascaded MRT collision step. It reduces to a corresponding TRT collision step, when $\omega^{\mathrm{even}}=\omega_4=\omega_6=\omega_8$ and $\omega^{\mathrm{odd}}=\omega_3=\omega_5=\omega_7$. Also, a different perspective of the
generalized equilibrium, Eq.~(\ref{eq:generalizedequilibrium}) can be arrived at in light of the consistency analysis performed
in the previous section. For example, for the third-order moments, $\beta=6$ and $7$, Eq.~(\ref{eq:generalizedequilibrium}) needs the
non-equilibrium moments $\widehat{f}_3^{(neq)}$, $\widehat{f}_4^{(neq)}$ and $\widehat{f}_5^{(neq)}$, which can be approximated
by Eqs.~(\ref{eq:noneqmfhat3new}), (\ref{eq:noneqmfhat4new}) and (\ref{eq:noneqmfhat5new}), respectively, which actually provide expressions for the components of the strain rate tensor in the cascaded formulation. Thus, we get
\begin{eqnarray}
\widehat{f}_{6}^{eq,G}&\approx&\widehat{f}_{6}^{eq}-\frac{1}{3}\left(\frac{1}{\omega_3}-\frac{1}{\omega_6}\right)u_y\overrightarrow{\nabla}\cdot\overrightarrow{j}
-\frac{1}{3}\left(\frac{1}{\omega_4}-\frac{1}{\omega_6}\right)u_y\left(\partial_xj_y-\partial_yj_x\right)
\nonumber\\
&&-\frac{2}{3}\left(\frac{1}{\omega_5}-\frac{1}{\omega_6}\right)u_x\left(\partial_xj_y+\partial_yj_x\right),\\
\widehat{f}_{7}^{eq,G}&\approx&\widehat{f}_{7}^{eq}-\frac{1}{3}\left(\frac{1}{\omega_3}-\frac{1}{\omega_7}\right)u_x\overrightarrow{\nabla}\cdot\overrightarrow{j}
-\frac{1}{3}\left(\frac{1}{\omega_4}-\frac{1}{\omega_7}\right)u_x\left(\partial_xj_y-\partial_yj_x\right)
\nonumber\\
&&-\frac{2}{3}\left(\frac{1}{\omega_5}-\frac{1}{\omega_7}\right)u_y\left(\partial_xj_y+\partial_yj_x\right).
\end{eqnarray}
In other words, the generalized equilibrium is a function of density and velocity fields and their gradients, the coefficients
of the latter terms are given as difference of relaxation times of moments of different order.

\section{\label{app:TimeImplicitnessCascadedCollision}Introducing Time-implicitness in the Cascaded Collision Operator}
Here, let us investigate the possibility of developing an executable LBE formulation where implicitness in time is introduced in the cascaded collision kernel, which could be useful in certain applications. In particular, we extend Eq.~(\ref{eq:cascadedLBE2}) such that the cascaded collision operator $\Omega_{{\alpha}(\overrightarrow{x},t)}^{c}$ is now treated to be semi-implicit in time:
\begin{eqnarray}
f_{\alpha}(\overrightarrow{x}+\overrightarrow{e}_{\alpha},t+1)=
f_{\alpha}(\overrightarrow{x},t)
&+&\frac{1}{2}\left[(\mathcal{K}\cdot\mathbf{\widehat{g}})_{{\alpha}(\overrightarrow{x},t)}
+(\mathcal{K}\cdot \mathbf{\widehat{g}})_{{\alpha}(\overrightarrow{x}+\overrightarrow{e}_{\alpha},t+1)}\right]\nonumber\\
&+&\frac{1}{2}\left[S_{{\alpha}(\overrightarrow{x},t)}+S_{{\alpha}(\overrightarrow{x}+\overrightarrow{e}_{\alpha},t+1)}\right]
\label{eq:cascadedLBE4}
\end{eqnarray}
In order to avoid an iterative procedure for the use of Eq.~(\ref{eq:cascadedLBE4}), we now define the following transformation with the introduction of a new variable $\overline{h}_{\alpha}$:
\begin{equation}
\overline{h}_{\alpha}=f_{\alpha}-\frac{1}{2}(\mathcal{K}\cdot\mathbf{\widehat{g}})_{\alpha}-\frac{1}{2}S_{\alpha}.
\label{eq:newtransformation}
\end{equation}
Now, substituting Eq.~(\ref{eq:newtransformation}) in Eq.~(\ref{eq:cascadedLBE4}), we get
\begin{equation}
\overline{h}_{\alpha}(\overrightarrow{x}+\overrightarrow{e}_{\alpha},t+1)-\overline{h}_{\alpha}(\overrightarrow{x},t)=
(\mathcal{K}\cdot\mathbf{\widehat{g}})_{{\alpha}(\overrightarrow{x},t)}+S_{{\alpha}(\overrightarrow{x},t)}
\label{eq:cascadedLBE5}
\end{equation}
As a result, Eq.~(\ref{eq:cascadedLBE5}) now becomes effectively explicit. In the new variable, the hydrodynamic fields can be obtained as $\rho=\sum_{\alpha=0}^{8}\overline{h}_{\alpha}$ and $\rho u_i=\sum_{\alpha=0}^{8}\overline{h}_{\alpha} e_{\alpha i}+\frac{1}{2}F_i$. The post-collision values, i.e. $\widetilde{\overline{h}}_{\alpha}$ can be obtained by replacing $\overline{f}_{\alpha}$ with $\overline{h}_{\alpha}$ in Eqs.~(\ref{eq:postcollisionf0})-(\ref{eq:postcollisionf8}). Now, to obtain the collision kernel $(\mathcal{K}\cdot\mathbf{\widehat{g}})_{\alpha}$ in Eq.~(\ref{eq:cascadedLBE5}) in terms of $\overline{h}_{\alpha}$, we define the following raw moment of
order $(m+n)$:
\begin{equation}
\widehat{\overline{\eta}}_{x^m y^n}^{'}=\sum_{\alpha}\overline{h}_{\alpha}e_{\alpha x}^m e_{\alpha y}^n
=\braket{e_{\alpha x}^m e_{\alpha y}^n|\overline{h}_{\alpha}},
\label{eq:newrawmoment}
\end{equation}
where $\widehat{\overline{\eta}}_{x^m y^n}^{'}$ can be represented and computed in a manner similar to that given in Eqs.~(\ref{eq:nonconservedbasisvector3})-(\ref{eq:nonconservedbasisvector8}).
From Eqs.~(\ref{eq:newtransformation}) and (\ref{eq:newrawmoment}), we obtain
\begin{eqnarray}
\widehat{\overline{\eta}}_{x^m y^n}^{'}&=&\widehat{\kappa}_{x^m y^n}^{'}
-\frac{1}{2}\sum_{\beta} \braket{K_{\beta}|e_{\alpha x}^m e_{\alpha y}^n}\widehat{g}_{\beta}
-\frac{1}{2}\widehat{\sigma}_{x^m y^n}^{'}\nonumber\\
&=&\widehat{\overline{\kappa}}_{x^m y^n}^{'}
-\frac{1}{2}\sum_{\beta} \braket{K_{\beta}|e_{\alpha x}^m e_{\alpha y}^n}\widehat{g}_{\beta}
\label{eq:newmomentrelation}
\end{eqnarray}
where $\sum_{\beta} \braket{K_{\beta}|e_{\alpha x}^m e_{\alpha y}^n}\widehat{g}_{\beta}$ can be obtained by exploiting the
orthogonal properties of $\mathcal{K}$, i.e. from Eqs.~(\ref{eq:collisionkernelmoment0})-(\ref{eq:collisionkernelmoment8}).

Now substituting Eq.~(\ref{eq:newmomentrelation}) in the collision kernel written in compact notation as given in Appendix~\ref{app:ChapmanEnskoganalysis}, i.e. in Eqs.~(\ref{eq:collisionkernelcompact3})-(\ref{eq:collisionkernelcompact8}), and simplifying we get
\begin{eqnarray}
\widehat{g}_3&=&\frac{1}{12}\frac{\omega_3}{\left(1+\frac{1}{2}\omega_3\right)}\left\{ \frac{2}{3}\rho+\rho(u_x^2+u_y^2)
-(\widehat{\overline{\eta}}_{xx}^{'}+\widehat{\overline{\eta}}_{yy}^{'})
-\frac{1}{2}(\widehat{\sigma}_{xx}^{'}+\widehat{\sigma}_{yy}^{'})
\right\}, \label{eq:newcollisionkernelcompact3}\\
\widehat{g}_4&=&\frac{1}{4}\frac{\omega_4}{\left(1+\frac{1}{2}\omega_4\right)}\left\{\rho(u_x^2-u_y^2)
-(\widehat{\overline{\eta}}_{xx}^{'}-\widehat{\overline{\eta}}_{yy}^{'})
-\frac{1}{2}(\widehat{\sigma}_{xx}^{'}-\widehat{\sigma}_{yy}^{'})
\right\}, \label{eq:newcollisionkernelcompact4}\\
\widehat{g}_5&=&\frac{1}{4}\frac{\omega_5}{\left(1+\frac{1}{2}\omega_5\right)}\left\{\rho u_x u_y
-\widehat{\overline{\eta}}_{xy}^{'}
-\frac{1}{2}\widehat{\sigma}_{xy}^{'}
\right\}, \label{eq:newcollisionkernelcompact5}\\
\widehat{g}_6&=&\frac{1}{4}\frac{\omega_6}{\left(1+\frac{1}{2}\omega_6\right)}\left\{2\rho u_x^2 u_y+\widehat{\overline{\eta}}_{xxy}^{'}
              -2u_x\widehat{\overline{\eta}}_{xy}^{'}-u_y\widehat{\overline{\eta}}_{xx}^{'}-\frac{1}{2}\widehat{\sigma}_{xxy}
              \right\}\nonumber\\
              &&-\frac{1}{2}u_y(3\widehat{g}_3+\widehat{g}_4)-2u_x\widehat{g}_5, \label{eq:newcollisionkernelcompact6}\\
\widehat{g}_7&=&\frac{1}{4}\frac{\omega_7}{\left(1+\frac{1}{2}\omega_7\right)}\left\{2\rho u_x u_y^2+\widehat{\overline{\eta}}_{xyy}^{'}
              -2u_y\widehat{\overline{\eta}}_{xy}^{'}-u_x\widehat{\overline{\eta}}_{yy}^{'}-\frac{1}{2}\widehat{\sigma}_{xyy}
              \right\}\nonumber\\
              &&-\frac{1}{2}u_x(3\widehat{g}_3-\widehat{g}_4)-2u_y\widehat{g}_5, \label{eq:newcollisionkernelcompact7}\\
\widehat{g}_8&=&\frac{1}{4}\frac{\omega_8}{\left(1+\frac{1}{2}\omega_8\right)}\left\{\frac{1}{9}\rho+3\rho u_x^2 u_y^2-\left[\widehat{\overline{\eta}}_{xxyy}^{'}
                                 -2u_x\widehat{\overline{\eta}}_{xyy}^{'}-2u_y\widehat{\overline{\eta}}_{xxy}^{'}
                                 +u_x^2\widehat{\overline{\eta}}_{yy}^{'}+u_y^2\widehat{\overline{\eta}}_{xx}^{'}\right.\right.\nonumber\\
                                 &&\left.\left.
                                 +4u_xu_y\widehat{\overline{\eta}}_{xy}^{'}
                                 \right]-\frac{1}{2}\widehat{\sigma}_{xxyy}^{'}\right\}
                                 -2\widehat{g}_3-\frac{1}{2}u_y^2(3\widehat{g}_3+\widehat{g}_4)
                                  -\frac{1}{2}u_x^2(3\widehat{g}_3-\widehat{g}_4)\nonumber\\
                                  &&-4u_xu_y\widehat{g}_5-2u_y\widehat{g}_6
                                  -2u_x\widehat{g}_7.\label{eq:newcollisionkernelcompact8}
\end{eqnarray}
It may be noted that a Chapman-Enskog analysis, as given in Appendix~\ref{app:ChapmanEnskoganalysis}, when performed with the above collision operator, yields the following relations between relaxation parameters and transport coefficients (see Eq.~(\ref{eq:coeffrelaxparameters})): $\vartheta_3=\frac{1}{3\omega_3},\quad \vartheta_4=\frac{1}{3\omega_4},\quad \vartheta_5=\frac{1}{3\omega_5}$, for the hydrodynamical equations given in Eq.~(\ref{eq:hydromoments1new}) and ~(\ref{eq:hydromoments2new}). Thus, the above considerations show that it is possible to introduce time-implicitness in the cascaded collision kernel, and when a transformation is introduced to make the computational procedure effectively explicit, it leaves the form of $\widehat{g}_{\beta}$ unchanged with a simple re-scaling of the relaxation parameters.

\newpage 

\end{document}